\documentclass{ws-ijgmmp}
\usepackage{amsmath}
\usepackage{amssymb}
\numberwithin{equation}{section}

\begin{document}
\markboth{E. Battista, G. Esposito, P. Scudellaro, F. Tramontano}
{Riemann curvature of a boosted spacetime geometry}

\title{RIEMANN CURVATURE OF A BOOSTED SPACETIME GEOMETRY}

\author{EMMANUELE BATTISTA}
\address{Dipartimento di Fisica, Complesso Universitario di 
Monte S. Angelo, Via Cintia Edificio 6, 80126 Napoli, Italy\\
Istituto Nazionale di Fisica Nucleare, Sezione di
Napoli, Complesso Universitario di Monte S. Angelo, 
Via Cintia Edificio 6, 80126 Napoli, Italy\\
\email{ebattista@na.infn.it}}

\author{GIAMPIERO ESPOSITO}
\address{Istituto Nazionale di Fisica Nucleare, Sezione di
Napoli, Complesso Universitario di Monte S. Angelo, 
Via Cintia Edificio 6, 80126 Napoli, Italy\\
\email{gesposit@na.infn.it}}

\author{PAOLO SCUDELLARO}
\address{Dipartimento di Fisica, Complesso Universitario di 
Monte S. Angelo, Via Cintia Edificio 6, 80126 Napoli, Italy\\
Istituto Nazionale di Fisica Nucleare, Sezione di
Napoli, Complesso Universitario di Monte S. Angelo, 
Via Cintia Edificio 6, 80126 Napoli, Italy\\
\email{scud@na.infn.it}}

\author{FRANCESCO TRAMONTANO}
\address{Dipartimento di Fisica, Complesso Universitario di 
Monte S. Angelo, Via Cintia Edificio 6, 80126 Napoli, Italy\\
Istituto Nazionale di Fisica Nucleare, Sezione di
Napoli, Complesso Universitario di Monte S. Angelo, 
Via Cintia Edificio 6, 80126 Napoli, Italy\\
\email{tramonta@na.infn.it}}

\maketitle

\begin{abstract}
The ultrarelativistic boosting procedure had been applied in the
literature to map the metric of Schwarzschild-de Sitter spacetime
into a metric describing de Sitter spacetime plus a shock-wave
singularity located on a null hypersurface. This paper evaluates the
Riemann curvature tensor of the boosted Schwarzschild-de Sitter metric  by means of numerical calculations, which make it possible to 
reach the ultrarelativistic regime gradually by letting the boost velocity approach the speed of light. Thus, for the first time in the 
literature, the singular limit of curvature, through Dirac's $\delta$ distribution and its derivatives, 
is numerically evaluated for this class of spacetimes. 
Moreover, the analysis of the Kretschmann invariant and the geodesic equation shows that the spacetime possesses a 
``scalar curvature singularity'' within a 3-sphere and it is possible to define what we here call ``boosted horizon'', a sort of elastic 
wall where all particles are surprisingly pushed away, as numerical analysis demonstrates. This seems to suggest that such 
``boosted geometries'' are ruled by a sort of ``antigravity effect'' since all geodesics seem to refuse to enter the ``boosted horizon'' 
and are ``reflected'' by it, even though their initial conditions are aimed at driving the 
particles towards the ``boosted horizon'' itself. Eventually, the equivalence with the coordinate 
shift method is invoked in order to demonstrate that all $\delta^2$ terms appearing in the Riemann 
curvature tensor give vanishing contribution in distributional sense.
\end{abstract}

\keywords{boost, black hole, singularity}

\section{Introduction}

The subject of gravitational fields generated by sources which move at the speed of light has been extensively studied in the literature 
because of its close connection to the topic of gravitational waves\footnote{In particular we talk about gravitational shock-waves.}, 
whose direct detection remains extremely difficult, since one normally 
deals with a very weak signal. The first who dealt with this aspect of general relativity was Tolman in 1934 \cite{Tolman1934}, who studied the 
gravitational field of light beams and pulses in the linearized theory. But it was only in 1971 that Aichelburg and Sexl  \cite{AS1971} 
developed a method to describe the gravitational field associated to a massless point particle moving at the 
speed of light in Minkowski spacetime (i.e. the gravitational 
field from a single photon). In fact  in Ref. \cite{AS1971} the authors first derive this field by solving the linearized Einstein field equations 
for a particle with rest mass $m$ moving uniformly with velocity $v$. Then they take the limit $v \rightarrow 1$ while the mass of the particle 
tends to zero in such a way that its energy remains finite. After that, they start with the full Einstein 
theory and the Schwarzschild metric\footnote{For a modern and innovative review of Schwarzschild solution see Ref. \cite{Antoci}.} 
(the exact metric describing a particle at rest), which written in isotropic coordinates reads as
\begin{equation}
{\rm d}s^2 = \dfrac{(1-A)^2}{(1+A)^2}{\rm d}t^2 -(1+A)^4 ({\rm d}x^2+{\rm d}y^2+{\rm d}z^2) \label{isotropic_Schwarzchild},
\end{equation}
with $A=m/2r$ and $r^2 = x^2 + y^2 + z^2$. Afterwards they apply to this metric a Lorentz transformation 
\begin{equation}
\bar{t}  =  (1 - v^2)^{-1/2} (t+vx), \label{ASLorentztransformation1}
\end{equation}
\begin{equation}
\bar{x}  =  (1 - v^2)^{-1/2} (x+vt), 
\end{equation}
\begin{equation}
\bar{y}  =  y,
\end{equation}
\begin{equation}
\bar{z}  =  z \label{ASLorentztransformation2},
\end{equation}
to obtain the gravitational field as seen by an observer moving uniformly with velocity $v$ relative to the mass. Once the limits 
$v \rightarrow 1$ and $m \rightarrow 0$ are taken, Aichelburg and Sexl obtain the remarkable result that both the linearized solution and the 
exact solution agree completely. 

The method first developed by Aichelburg and Sexl is called in the literature ``the boost of a metric''. With this procedure it is possible to 
show that the gravitational field of a null source is nonvanishing on a plane containing the particle and orthogonal to the direction of 
motion, i.e. (asymmetric) plane-fronted gravitational waves. The Riemann curvature tensor is zero everywhere except on this plane, where it assumes a 
distributional nature. The intriguing fact is that the boosted metric in the ultrarelativistic regime ($v \rightarrow 1$) has a new type of 
singularity, i.e. a distributional (Dirac-delta-like) singularity. The boosted ultrarelativistic metric obtained in Ref. \cite{AS1971} reads 
indeed as
\begin{equation}
{\rm d}s^2 = {\rm d}\bar{t \;}^2 - {\rm d}\bar{x}^2  - {\rm d}\bar{y}^2  - {\rm d}\bar{z}^2 
- 4 p \{ (\lvert \bar{t}-\bar{x} \rvert)^{-1} -2 \delta (\bar{t \;}^2 - \bar{x}^2) \log\sqrt{\bar{y} +\bar{z}} \;  \} 
({\rm d}\bar{t} - {\rm d}\bar{x})^2, \label{ASmetric}
\end{equation}
with $p \equiv m / \sqrt{1-v^2}$. Thus, the gravitational field turns out to travel with the particle, 
being zero everywhere except at the hypersurface $\bar{t}=\bar{x}$. Moreover, as anticipated before, the Riemann 
tensor of (\ref{ASmetric}) is zero everywhere except on the hypersurface $\bar{t}=\bar{x}$ and has nonvanishing components given by \cite{AS1971}
\begin{equation}
R_{0202}=4p \; \delta (\bar{t}-\bar{x}) \left[\dfrac{\bar{y}^2-\bar{z}^2}{(\bar{y}^2+\bar{z}^2)^2} 
+ \pi \delta(\bar{y})\delta(\bar{z})\right], \label{ASRiemann1}
\end{equation}
\begin{equation}
R_{0303}=4p \; \delta (\bar{t}-\bar{x}) \left[\dfrac{\bar{y}^2-\bar{z}^2}{(\bar{y}^2+\bar{z}^2)^2} - \pi \delta(\bar{y})\delta(\bar{z})\right],
\end{equation}
\begin{equation}
R_{0203}=-4p \; \delta (\bar{t}-\bar{x})\dfrac{2\bar{y}\bar{z}}{(\bar{y}^2+\bar{z}^2)^2},\label{ASRiemann2}
\end{equation}
with the other components related to the ones given above by symmetry. An important remark should be made at this point, 
since the Riemann tensor is perfectly defined as it contains the tensor product of Dirac's $\delta$ distributions 
(and not their multiplications). The only elements which are ``poorly defined'' in 
(\ref{ASRiemann1})--(\ref{ASRiemann2}) are represented by the functions
\begin{equation}
\dfrac{\bar{y}^2-\bar{z}^2}{(\bar{y}^2+\bar{z}^2)^2},
\end{equation}
and
\begin{equation}
\dfrac{2\bar{y}\bar{z}}{(\bar{y}^2+\bar{z}^2)^2},
\end{equation}
which are not locally integrable on the $(y,z)$-plane and, therefore, do not define, a priori, any distribution. 
Of course, their ``regularization'' (a la Gel'fand, see for example Ref. \cite{Gelfand}) is straightforward: the integration 
is understood in such a way that we first integrate over the set $y^2 + z^2 > \epsilon$ and then pass to the limit $\epsilon \rightarrow 0$.

In order to give a precise meaning to expressions (\ref{ASRiemann1})--(\ref{ASRiemann2}), instead of transforming the 
Schwarzschild metric (\ref{isotropic_Schwarzchild}), the authors of Ref. \cite{AS1971} have applied the Lorentz 
transformations (\ref{ASLorentztransformation1})--(\ref{ASLorentztransformation2}) directly to the components of the 
Riemann tensor and then they have investigated the regime $v \rightarrow 1$. In this way, with the help of tetrad formalism, 
Aichelburg and Sexl have obtained relations which are valid only for those spacetime points where $\bar{y}^2+\bar{z}^2 \neq 0$. 
In particular, they obtain again the relations (\ref{ASRiemann1})--(\ref{ASRiemann2}), but without the functions 
$\delta(\bar{y})\delta(\bar{z})$ which vanish because of the condition $\bar{y}^2+\bar{z}^2 \neq 0$. This fact shows that on 
the hypersurface $\bar{x}=\bar{t}$ the Riemann tensor has a $\delta$-like singularity and is exactly of Petrov type $N$ 
(i.e. all four principal null directions of the Weyl spinor, describing the Weyl conformal curvature, coincide). 

Years after the work by Aichelburg and Sexl, more general impulsive waves were obtained by boosting other
black hole spacetimes with rotation, charge and a cosmological constant 
\cite{BN1995,FP1990,LS1992,HT1993,BN1995,BN1996,PG1997,PG1998,Ortaggio,ES2007}. 
The technique of boosting a spacetime metric in fact has a lot of applications in theoretical physics. The work in Ref. \cite{Ar2009a}, 
for instance, shows that the black hole formation caused by the collision of two particles with large relative velocity ($v \rightarrow 1$), 
and considered in the rest frame of one of the particles, involves the concept of boosted metric: the gravitational field of the other particle 
is described by the ultrarelativistic boosted Schwarschild-de Sitter metric. Moreover, the collisions of shock-waves and heavy ions as well as 
the entropy that is consequently produced \cite{GPY2008} appeal to the boost procedure, also in the context of higher dimensions \cite{Tal2012} 
and branes \cite{AnSo2007,Sfetsos2005,Kaloper2005,Ar2000}. Furthermore, it is possible to study the formation of marginally trapped surfaces 
in the head-on collision both of two shock-waves \cite{Ar2009b} and of two ultrarelativistic charged particles \cite{Ar2010} in de Sitter space 
by using the procedure of boosting a metric, since for example in the latter case the metric of the two charges is obtained by boosting 
Reissner-Nordstr\"om-de Sitter spacetime to the speed of light, while with similar arguments it is shown in Ref. \cite{VV2012} that the 
collision of two Reissner-Nordstr\"om gravitational shock-waves in anti-de Sitter space prevents the formation of marginally trapped surfaces 
of Penrose type. Finally, the concept of a boosted metric can be used as a tool to describe (de Sitter) spacetime from a quantum point of 
view \cite{GM2007}.

Our main attention here will be devoted to the work in Refs. \cite{HT1993}, \cite{ES2007} where it has been shown in detail how to map, 
through a boosting procedure, the Schwarzschild-de Sitter metric
\begin{equation}
{\rm d}s^{2}=-\left(1-{2m \over r}-{r^{2}\over a^{2}}\right){\rm d}t^{2}
+{{\rm d}r^{2}\over \left(1-{2m \over r}-{r^{2}\over a^{2}}\right)}
+r^{2}({\rm d}\theta^{2}+\sin^{2}\theta \;  {\rm d}\phi^{2}) , \label{S-dS metric}
\end{equation}
into the highly singular form\footnote{For the manifestly four-dimensional form of metric 
(\ref{ultrarelativistic boosted metric}) see Appendix \ref{Tetrad}.} (with $v \rightarrow 1$) 
\begin{equation}
\begin{split}
{\rm d}s^{2}=&-{\rm d}Y_{0}^{2}+{\rm d}Y_{1}^{2}+{\rm d}Y_{2}^{2}+{\rm d}Y_{3}^{2}+{\rm d}Y_{4}^{2}  \\
&+ 4 p \left[-2+{Y_{4}\over a}
\log \left({{a+Y_{4}}\over {a-Y_{4}}}\right)\right]
\delta(Y_{0}+Y_{1})({\rm d}Y_{0}+{\rm d}Y_{1})^{2}, \label{ultrarelativistic boosted metric}
\end{split}
\end{equation}
where the first line describes de Sitter space viewed as a four-dimensional hyperboloid of radius $a$ having equation
\begin{equation}
(Y_{0})^{2}=-a^{2}+(Y_{1})^{2}+(Y_{2})^{2}+(Y_{3})^{2}+(Y_{4})^{2},  \label{hyperboloid constrain}
\end{equation}
embedded into flat five-dimensional space, while the second line
of (\ref{ultrarelativistic boosted metric}) describes a shock-wave singularity located on the null
hypersurface having equations
\begin{equation}
Y_{0}+Y_{1}=0, \label{null hypersurface 1}
\end{equation}
\begin{equation}
(Y_{2})^{2}+(Y_{3})^{2}+(Y_{4})^{2}-a^{2}=0, \label{null hypersurface 2}
\end{equation}
equation (\ref{null hypersurface 2}) being obtained by the joint effect of the hyperboloid
constraint (\ref{hyperboloid constrain}) and the Dirac-delta condition (\ref{null hypersurface 1}). Since the 
metric is turned into a mathematical object having distributional
nature, the usual spacetime picture is no longer valid, but it would
be very interesting to evaluate the effect of these shock-wave
singularities on curvature. The great revolution introduced by Einstein's theory consists in fact in viewing the gravitational field as the 
curvature of spacetime. Such a curvature is directly coupled to the energy and momentum of whatever matter and radiation are present, as 
specified by the Einstein field equations whose content states that ``the matter and the energy say to the spacetime how to curve, and the 
curvature of spacetime says to the matter how to move'' \cite{MTW}. Thus, one of the most important objects of the theory of the gravitational 
field is the Riemann tensor, since it is an intrinsic object that catches in an elegant and covariant way the features of spacetime curvature. 
Therefore it could be of great physical importance to evaluate the Riemann tensor for this type of geometries, i.e.  
``the boosted geometries''. (To fully appreciate the importance of this tensor see Appendix \ref{Appendix Riemann}). 

Since ``gravitation is a manifestation of spacetime curvature, and curvature shows up in the deviation of one geodesic from a nearby 
geodesic'' \cite{MTW}, the concept of spacetime curvature is directly related to the geodesic completeness of spacetime, as we say that 
a spacetime manifold is geodesically complete if any geodesic can be extended to arbitrary values of the affine parameter. Thus, knowledge of 
the Riemann curvature tensor is an essential step towards the description of topological features of spacetime and this motivates the effort 
we made in calculating the Riemann tensor for the boosted Schwarschild-de Sitter metric. 

We stress that the definitions (\ref{definition Riemann})--(\ref{geodesic deviation}) are given in terms of objects that, unlike the ones 
we will handle, have no distributional singularities (cfr (\ref{ultrarelativistic boosted metric})). Thus, in this article we are interested 
in a sort of generalization of the usual concept of Riemann tensor, which enlarges the notion of curvature, i.e. what we call the ``boosted 
Riemann tensor'', with a particular interest in the ultrarelativistic regime, where distributional singularities show up. By virtue of the high 
difficulty of dealing with the metric (\ref{ultrarelativistic boosted metric}), we decided to start from its low-velocity limit and then to 
reach the ultrarelativistic regime via numerical calculations. For this purpose, Sec. II evaluates the procedure to obtain the boosted 
Schwarzschild-de Sitter metric in manifestly four-dimensional form. Then, it is shown that the basis defined by the boost procedure is a 
coordinate basis, a property that greatly simplifies the calculations performed. In Sec. III  the analysis of both the Kretschmann invariant 
and the geodesic equation allows us  to characterize the features of curvature. 

An important question arises while dealing with Secs. II-III, i.e. how to deal with the Riemann curvature tensor when 
it has terms proportional to $\delta^2$. In fact from (\ref{ultrarelativistic boosted metric}) it is easy to understand that 
the Riemann tensor has got terms involving the products of two Dirac's $\delta$ distributions (a formal method to cope with 
multiplication of distributions can be found in Ref. \cite{Colombeau}). This means that the ``boosted Riemann tensor'' of the 
``boosted geometry'' we are going to describe is in principle not defined. Anyway, we will be able to show that the $\delta^2$ 
terms appearing in the ``boosted Riemann tensor'' vanish in a distributional sense. Unlike the (rather simple) example 
discussed in Ref. \cite{AS1971}, we will achieve this point in a more difficult way, since the high difficulty of metric 
(\ref{ultrarelativistic boosted metric}) makes it quite impossible to write down explicitly all the boosted Riemann tensor 
components, as anticipated above. For this reason in Sec. IV we will make use of an equivalent method to describe the gravitational 
shock-wave of a massless particle, i.e. the coordinate shift method \cite{Dray,Sfetsos95} (or the scissors-and-paste method by 
Penrose \cite{Penrose72}). The equivalence of this method and the boosting procedure has been demonstrated by the authors of 
Ref. \cite{Dray}, where it is explicitly shown that with the new approach it is possible to recover the results of Aichelburg and 
Sexl. By exploiting this equivalence between the two methods, we will be able to show in which sense the $\delta^2$ terms 
appearing in the Riemann tensor of metric (\ref{ultrarelativistic boosted metric}) can be seen as vanishing, leading to a well 
defined spacetime function (in the sense of distributions). Concluding remarks and open problems are presented in Sec. V.

\section{The ``boosted'' Riemann curvature tensor}

Following Refs. \cite{HT1993}, \cite{ES2007} we can express a de Sitter spacetime in four dimensions as a four-dimensional hyperboloid 
of radius $a$ embedded in five-dimensional Minkowski spacetime having metric
\begin{equation}
{\rm d}s^{2}_{M}= - {\rm d}Z^{2}_{0}+ {\rm d}Z_{1}^{2}+{\rm d}Z_{2}^{2}+{\rm d}Z_{3}^{2}+{\rm d}Z_{4}^{2},
\end{equation}
with coordinates satisfying the hyperboloid constraint 
\begin{equation}
a^2 = -(Z_{0})^2+ (Z_{1})^{2}+(Z_{2})^{2}+(Z_{3})^{2}+(Z_{4})^{2}.
\label{Z hyperboloid constraint}
\end{equation}
By exploiting the relations between the $Z_i$ ($i=0,1,2,3,4$) coordinates and the spherical static coordinates $(t,r,\theta,\phi)$ 
\begin{equation}
Z_0 \equiv  \sqrt{a^2 - r^2} \sinh(t/a), \label{Z0}
\end{equation}
\begin{equation}
Z_1 \equiv  r \cos \theta, 
\end{equation}
\begin{equation}
Z_2 \equiv  r \sin \theta \cos \phi , 
\end{equation}
\begin{equation}
Z_3 \equiv  r \sin \theta \sin \phi,  
\end{equation}
\begin{equation}
Z_4 \equiv  \pm  \sqrt{a^2 - r^2} \cosh(t/a), \label{Z3}
\end{equation}
and on defining 
\begin{equation}
f^2 \equiv  a^2 - r^2 = (Z_{4})^2 - (Z_{0})^{2}, 
\end{equation}
\begin{equation}
F_m \equiv   1 - \dfrac{2 a^2 m}{f^2 r}-\dfrac{a^2/r^2}{\left( 1- \dfrac{2 a^2 m}{f^2 r}\right)}, 
\end{equation}
\begin{equation}
Q \equiv   1+ \dfrac{2 (Z_{0})^{2}}{f^2},
\end{equation}
we can express the Schwarschild-de Sitter metric (\ref{S-dS metric}) in the form 
\begin{equation}
{\rm d}s^2 = h_{00}{\rm d}Z_{0}^{2}+ h_{44}{\rm d}Z_{4}^{2}+2h_{04}{\rm d}Z_{0}{\rm d}Z_{4}+ {\rm d}Z_{1}^{2}
+{\rm d}Z_{2}^{2}+{\rm d}Z_{3}^{2},
\label{Z S-dS metric}
\end{equation} 
where
\begin{equation}
h_{00} \equiv -\dfrac{1}{2} \left(Q-1 \right)F_m - \left(1-\dfrac{2 a^2 m}{f^2 r} \right) - \dfrac{(Z_{0})^{2}}{r^2}, \\
\end{equation}
\begin{equation}
h_{44} \equiv  -\dfrac{1}{2} \left(Q+1 \right)F_m + \left(1-\dfrac{2 a^2 m}{f^2 r} \right) - \dfrac{(Z_{4})^{2}}{r^2}, \\
\end{equation}
\begin{equation}
h_{04} \equiv  \dfrac{Z_0 Z_4}{f^2} F_m+ \dfrac{Z_0 Z_4}{r^2}.  
\end{equation}
At this stage, we introduce a Lorentz boost in the $Z_1$-direction by defining a new set of coordinates independent of $v$, i.e. the $Y_i$ 
coordinates, such that (hereafter $\gamma \equiv  1/ \sqrt{1-v^2} \; $)
\begin{equation}
Z_0 = \gamma \left( Y_0 + v Y_1 \right),  \label{boost 1}
\end{equation}
\begin{equation}
Z_1 = \gamma \left( v Y_0 + Y_1 \right), 
\end{equation}
\begin{equation}
Z_2 = Y_2, \; \;  \; Z_3=Y_3 , \; \;  \; Z_4=Y_4. \label{boost 2}
\end{equation}
Thus, starting from (\ref{Z S-dS metric}) jointly with (\ref{boost 1})--(\ref{boost 2}) we eventually obtain the boosted 
Schwarzschild-de Sitter metric
\begin{equation}
\begin{split}
{\rm d}s^2 =& \gamma^2 \left(h_{00}+v^2\right){\rm d}Y_{0}^{2}+\gamma^2 \left(1+ v^2 h_{00}\right)
{\rm d}Y_{1}^{2}+{\rm d}Y_{2}^{2}+{\rm d}Y_{3}^{2}+h_{44}{\rm d}Y_{4}^{2}   \\
& + 2v\gamma^2 \left(1+h_{00}\right){\rm d}Y_0 {\rm d}Y_1 + 2 \gamma h_{04} {\rm d}Y_0 {\rm d}Y_4 
+ 2 v \gamma h_{04} {\rm d}Y_1 {\rm d}Y_4, \label{boosted metric}
\end{split}
\end{equation}
whose singular ultrarelativistic limit is expressed by (\ref{ultrarelativistic boosted metric}). Thus, we can interpret 
(\ref{boosted metric}) as the low-velocity limit of (\ref{ultrarelativistic boosted metric}). 

The spacetime metric (\ref{boosted metric}) is apparently expressed by a $5 \times 5$ matrix while the original metric 
(\ref{S-dS metric}) is expressed through $4$ local coordinates $t,r,\theta,\phi$. Hence also the metric  (\ref{boosted metric}) 
should be eventually expressed through $4$ coordinates only, if
one wants to arrive at a formula for the curvature, since our reference spacetime
remains four-dimensional. To restore the usual four-dimensional form of the metric, we have to exploit the constraint 
(\ref{Z hyperboloid constraint}) expressed in terms of $Y_i$ coordinates, i.e. Eq. (\ref{hyperboloid constrain}). By virtue of 
this condition we can write
\begin{equation}
Y_{0}=\sqrt{-a^{2}+(Y_{1})^{2}+(Y_{2})^{2}+(Y_{3})^{2}+(Y_{4})^{2}} \equiv \sqrt{\sigma(Y_{\mu })}, \label{Y0}
\end{equation}
\begin{equation}
dY_{0}={\sum_{\mu=1}^{4}Y_{\mu}dY_{\mu }\over \sqrt{\sigma(Y_{\mu })}}, \label{dYo}
\end{equation}
and eventually, using (\ref{Y0}) and (\ref{dYo}), we obtain the manifestly four-dimensional form of the boosted metric 
(\ref{boosted metric}), which can be expressed by the relations
\begin{equation}
g_{11} =  \dfrac{\gamma^2 \left(h_{00}+v^2\right)}{\sigma} (Y_{1})^{2}+\gamma^2 \left(1+v^2h_{00}\right)+\dfrac{2v\gamma^2  
\left(1+h_{00}\right)}{\sqrt{\sigma}} Y_1 \label{g11} , 
\end{equation}
\begin{equation}
g_{22} = \dfrac{\gamma^2 \left(h_{00}+v^2\right)}{\sigma} (Y_{2})^{2} +1, 
\end{equation}
\begin{equation}
g_{33} = \dfrac{\gamma^2 \left(h_{00}+v^2\right)}{\sigma} (Y_{3})^{2} +1, 
\end{equation}
\begin{equation}
g_{44} = \dfrac{\gamma^2 \left(h_{00}+v^2\right)}{\sigma} (Y_{4})^{2} +h_{44}+ \dfrac{2 \gamma h_{04}}{\sqrt{\sigma}} Y_4 , 
\end{equation}
\begin{equation}
g_{12} = \dfrac{\gamma^2 \left(h_{00}+v^2\right)}{\sigma} Y_{1}Y_{2} + \dfrac{v \gamma^2 \left(1+h_{00} \right)}{\sqrt{\sigma}}Y_2, 
\end{equation}
\begin{equation}
g_{13} = \dfrac{\gamma^2 \left(h_{00}+v^2\right)}{\sigma} Y_{1}Y_{3} + \dfrac{v \gamma^2 \left(1+h_{00} \right)}{\sqrt{\sigma}}Y_3, 
\end{equation}
\begin{equation}
g_{14} = \dfrac{\gamma^2 \left(h_{00}+v^2\right)}{\sigma} Y_{1}Y_{4} + \dfrac{v \gamma^2 \left(1+h_{00} \right)}{\sqrt{\sigma}}Y_4 
+ \dfrac{\gamma h_{04}}{\sqrt{\sigma}}+v \gamma h_{04}, 
\end{equation}
\begin{equation}
g_{23} = \dfrac{\gamma^2 \left(h_{00}+v^2\right)}{\sigma} Y_{2}Y_{3} , 
\end{equation}
\begin{equation}
g_{24} = \dfrac{\gamma^2 \left(h_{00}+v^2\right)}{\sigma} Y_{2}Y_{4} + \dfrac{\gamma h_{04}}{\sqrt{\sigma}} Y_2 , 
\end{equation}
\begin{equation}
g_{34} = \dfrac{\gamma^2 \left(h_{00}+v^2\right)}{\sigma} Y_{3}Y_{4} + \dfrac{\gamma h_{04}}{\sqrt{\sigma}} Y_3. \label{g34}
\end{equation}
Having obtained the formulas (\ref{g11})--(\ref{g34}), we can evaluate the Riemann-Christoffel symbols and consequently the Riemann 
curvature tensor of the boosted Schwarzschild-de Sitter metric by using the familiar relations of classical general relativity. 
The most general form of Riemann-Christoffel symbols reads as follows \cite{MTW} ($a,b,c$ being abstract indices):
\begin{equation}
\Gamma_{abc}= \dfrac{1}{2} \left( g_{ab,c}+g_{ac,b}-g_{bc,a}+c_{abc}+c_{acb}-c_{bca} \right), \label{Gamma}
\end{equation}
where the ``commutation coefficients''  $c_{abc}$ are defined by
\begin{equation}
[\bold{e}_{b},\bold{e}_{c}] \equiv c_{bc}^{\; \; \; a} \; \bold{e}_{a},  \label{c_abc}
\end{equation}
with $\{ \bold{e}_{a}\}$ being any noncoordinate basis. Last, the components of the Riemann tensor are given by
\begin{equation}
R^{a}_{\; \; bcd} = \Gamma^{a}_{\; \;  bd,c} - \Gamma^{a}_{\; \;  bc,d} + \Gamma^{e}_{\; \;  bd} \Gamma^{a}_{\; \;  ec}
-\Gamma^{e}_{\; \;  bc} \Gamma^{a}_{\; \;  ed}- \Gamma^{a}_{\; \;  be}c_{cd}^{\; \; \; e}. \label{Riemann tensor} 
\end{equation}
We can somewhat simplify the relations (\ref{Gamma}), (\ref{Riemann tensor}) in the case in which $ \left \{ 
\dfrac{\partial}{\partial Y_{\mu}} \right \}$ ($\mu$ being a coordinate index such that $\mu=1,2,3,4$) is a coordinate basis. As we know, 
the static spherical basis $(t,r,\theta,\phi)$ is indeed a coordinate basis. Bearing in mind definitions (\ref{Z0})--(\ref{Z3}), the 
Jacobian of the transformation between the spherical coordinates and the   $ \left \{ \dfrac{\partial}{\partial Z_{\mu}} \right \}$ 
is expressed by
\begin{equation}
J_{\mu} ^{\; \; \lambda} =
\begin{pmatrix}
0 & \cos \theta & -r \sin \theta & 0 \\
0 & \sin \theta \cos \phi &  r \cos \theta \cos \phi &  -r \sin \theta \sin \phi \\
0  & \sin \theta \sin \phi  & r \cos \theta \sin \phi & r \sin \theta \cos \phi  \\
\dfrac{\sqrt{a^2-r^2}}{a} \sinh (t/a) &  \dfrac{-r}{\sqrt{a^2-r^2} }\cosh (t/a) & 0 & 0 
\end{pmatrix} , 
\label{Jacobian}
\end{equation}
while the inverse Jacobian reads as
\begin{equation}
\begin{split}
& (J^{-1})_{\lambda}^{\; \; \mu} = \\
& \begin{pmatrix}
\dfrac{a \; r \cos \theta \coth (t/a)}{(a^2 - r^2)} &   \dfrac{a \; r \cos \phi  \sin \theta \coth (t/a)}{(a^2 - r^2)} 
&  \dfrac{a \; r \sin \theta \sin \phi \coth (t/a)}{(a^2 - r^2)}  & \dfrac{a \left(\sinh (t/a)\right)^{-1}}{\sqrt{a^2-r^2}} \\
\cos \theta & \cos \phi \sin \theta & \sin \theta \sin \phi & 0 \\
-\sin \theta /r & \cos \theta \cos \phi /r &   \cos \theta \sin \phi /r &  0\\
0  & - \dfrac{\sin \phi}{r \sin \theta} & \dfrac{\cos \phi}{r \sin \theta}&  0
\end{pmatrix} . 
\label{inverse jacobian}
\end{split}
\end{equation}
By virtue of (\ref{Jacobian}) and (\ref{inverse jacobian}), if we adopt the concise notation $x_{\lambda} \equiv 
(t,r,\theta,\phi)$ we can write 
\begin{equation}
\dfrac{\partial }{\partial Z_{\mu}}= (J^{-1})_{\lambda}^{\; \; \mu}  \dfrac{\partial }{\partial x_{\lambda}},
\end{equation}
and, by exploiting the fact that $ \left \{ \dfrac{\partial}{\partial x_{\lambda}} \right \}$ is a coordinate basis, after a 
lengthy calculation we arrive at the conclusion that also the basis $ \left \{ \dfrac{\partial}{\partial Z_{\mu}} \right \}$ is a 
coordinate basis, or in other words we have that 
\begin{equation}
\left[ \dfrac{\partial}{\partial Z_{\mu}} , \dfrac{\partial}{\partial Z_{\lambda}} \right]=0 . 
\end{equation}
The relations (\ref{boost 1})--(\ref{boost 2}) for the boost show that the transformations between $Z_{\mu}$ and $ Y_{\mu}$ are linear, 
therefore we can easily conclude that 
\begin{equation}
\left[ \dfrac{\partial}{\partial Y_{\mu}} , \dfrac{\partial}{\partial Y_{\lambda}} \right]=0 ,
\end{equation}
hence the basis   $ \left \{ \dfrac{\partial}{\partial Y_{\mu}} \right \}$ is a coordinate basis as well. 

This means that we can evaluate the Riemann-Christoffel symbols and the Riemann curvature tensor for the boosted spacetime metric 
(\ref{g11})--(\ref{g34}) by setting $c_{abc}=0$ in the relations (\ref{Gamma}), (\ref{Riemann tensor}). Nevertheless, these relations are 
still too complicated to be computed analytically, and therefore a numerical calculation has been necessary. Formulas 
(\ref{g11})--(\ref{g34}) show indeed that we are dealing with a spacetime metric represented by a $4 \times 4$ matrix whose elements 
are given by some complicated nonvanishing functions of the $Y_{\mu}$ coordinates. That is why we first tried to compute the Riemann 
curvature tensor analytically in terms of tetrads (see Appendix \ref{Tetrad}) before realizing that even this solution was far too 
complicated. Thus, the only way we had to compute the Riemann-Christoffel symbols and the Riemann tensor was represented by numerical 
calculations. In this way we can evaluate the behavior of spacetime curvature also in the ultra-relativistic regime, which is the one 
we are mainly interested in, by letting the velocity defined by the boost relations (\ref{boost 1})--(\ref{boost 2}) approach gradually 
the speed of light. 

In what follows we discuss the results of our computation mainly by studying curvature invariants and the behavior of geodesics in our 
reference spacetime. We in fact think that these features represent the best tools to describe physically the concept of spacetime curvature. 

\section{The Kretschmann invariant}

Intuitively, a spacetime singularity is a ``place'' where the curvature ``blows up'' \cite{Wald} or, by analogy with electrodynamics, 
a point where the metric tensor is either not defined or not suitably differentiable \cite{Haw-Ell}. Regrettably, both these statements 
are not rigorous definitions that can characterize the concept of singularity. First of all, since in general relativity we do not know the 
manifold and the metric structure in advance (they are solutions of Einstein field equations), we are not able to give a physical sense to the 
notion of an event until we solve Einstein equations, and hence the idea of a singularity as a ``place'' has not a satisfactory meaning. 
Moreover, also the notion of curvature becoming larger and larger as a general criterion for singularities has pathological problems. In fact, 
the bad behavior of components or derivatives of the Riemann tensor could be ascribed to the coordinate or tetrad basis used. To avoid this 
problem, one might examine scalar curvature invariants constructed from the Riemann tensor or its covariant derivatives, which in some cases 
can completely characterize the spacetime (see Refs. \cite{CHP,CBCR} for further details). However, even if the value of some scalar invariants 
is unbounded, curvature might blow up only ``as one goes to infinity'', a case that we would interpret as a singularity-free spacetime 
\cite{Wald}. Furthermore, spacetimes may be singular without any bad behavior of the curvature tensor (the so-called ``conical singularities'' 
\cite{Wald}). Lastly, the bad behavior of the metric tensor at some spacetime points cannot be a way to define singularities, as one could 
always cut out such points and hence the remaining manifold, representing the whole spacetime, would turn out to be nonsingular. 

A more satisfactory idea to define singularities is to use the notion of incompleteness of timelike geodesics, i.e. geodesics which are 
inextendible in at least one direction and hence have only a finite range of affine parameter. This has the immediate physical interpretation 
that there exist freely moving observers or particles whose histories did not exist after (or before) a finite interval of proper time. 
Although the physical meaning of affine parameter on null geodesics is different from the case of timelike geodesics, we could also regard 
null geodesic incompleteness as a good criterion to define spacetime singularities.  Thus, timelike and null geodesic completeness are minimum 
conditions for spacetime to be considered singularity-free \cite{Haw-Ell}. However, as there are examples of geodesically complete spacetimes 
which contain an inextendible timelike curve of bounded acceleration and finite length \cite{Geroch}, we should generalize the concept of 
affine parameter to all $C^1$ curves, no matter whether they are geodesics or not. This fact is linked to the concept of b-completeness 
(short for bundle completeness), which we shortly describe following Refs. \cite{Haw-Ell,Sch} in Appendix \ref{b-completeness}. 

Therefore, we can classify a singularity represented by the presence of at least one incomplete geodesic according to whether \cite{Wald}
\begin{enumerate}
\item a curvature invariant blows up along a geodesic (``scalar curvature singularity''),
\item a component of the Riemann tensor or its covariant derivatives in a parallelly propagated tetrad blows up along a geodesic 
(``parallelly propagated curvature singularity''),
\item no such invariant or component blows up (``noncurvature singularity'').
\end{enumerate}
We can therefore understand the importance of scalar curvature invariants in the analysis of spacetime singularities. Being coordinate 
independent, curvature invariants can describe the size of curvature and its growth along timelike curves, and can also characterize 
curvature singularities \cite{Thorpe}, while providing important information about the nature of singularities. For example, in the case of 
Schwarzschild metric, which can be obtained from (\ref{S-dS metric}) if we put $a=\infty$ (for an unambiguous definition of the notion of limit 
applied to spacetimes see Ref. \cite{Geroch69}), the Kretschmann invariant (i.e. the Riemann tensor squared) is such that 
\begin{equation}
R^{\alpha \beta \gamma \delta} R_{\alpha \beta \gamma \delta} = 48m^2 /r^6, \label{Schwarzschild kretsch}
\end{equation}
in agreement with the fact that in all coordinate systems 
the real singularity is located only at $r=0$ and not also at $r=2M$ (i.e. the event horizon). 

In order to study the features of the Riemann curvature of spacetime described by the metric (\ref{ultrarelativistic boosted metric}), 
we therefore decided to plot the Kretschmann invariant at different values of boost velocity $v$ and study the geodesic equation
\begin{equation}
\ddot{Y}^{\mu}(s) + \Gamma^{\mu}_{\; \nu \lambda} \dot{Y}^{\nu}(s) \dot{Y}^{\lambda}(s)=0, \label{geodesic eq}
\end{equation} 
$s$ being the affine parameter of the geodesic having parametric equation $Y^{\mu}=Y^{\mu}(s)$.

From the analysis of the Kretschmann invariant we found that it is not defined unless the inequality 
(hereafter, numerical values of $Y$ coordinates have downstairs indices, to be consistent with the notation in Sec. II)
\begin{equation}
(Y_{1})^2+(Y_{2})^2+(Y_{3})^2+(Y_{4})^2 > a^2, \label{singularity kretschmann}
\end{equation}
is satisfied. Hence, we see that the hyperboloid constraint, condition (\ref{hyperboloid constrain}), allows us to define a 3-sphere of 
radius $a$ where the Kretschmann invariant  is not defined. This peculiar feature of our ``boosted spacetime geometry'' is indeed obvious 
if we look at formulas (\ref{g11})--(\ref{g34}), as here the quantities $\sigma$ and $\sqrt{\sigma}$ always appear at the denominator of 
the expressions of the metric tensor $g_{\mu \nu}$, which means that the metric is defined only if the inequality 
(\ref{singularity kretschmann}) holds. Moreover, it is possible to derive Eq. (\ref{singularity kretschmann}) 
in  the regime $v<1$ from the analysis of the Kretschmann invariant for the Schwarzschild-de Sitter metric 
(\ref{S-dS metric}). In fact for (\ref{S-dS metric}) the Kretschmann invariant reads as
\begin{equation}
R^{\alpha \beta \gamma \delta} R_{\alpha \beta \gamma \delta} = 24 \left( \dfrac{1}{a^4}+\dfrac{2m^2}{r^6} \right), \label{S-dS kretsch}
\end{equation}
which reduces to (\ref{Schwarzschild kretsch}) in the limit $a=\infty$. Therefore, it follows immediately 
from (\ref{S-dS kretsch}) that the Schwarzschild-de Sitter metric (\ref{S-dS metric}) has an unique singularity located 
at $r=0$. Equation (\ref{r^2_1}) clearly shows that the condition $r=0$ leads to
\begin{equation}
\sqrt{\gamma^2 (v \sqrt{\sigma}+Y_1)^2+(Y_2)^2+(Y_3)^2}=0,
\end{equation}
which, being defined by the sum of squared quantities, in turns implies that
\begin{equation}
\left \{
\begin{array}{lll}
& v \sqrt{\sigma}+Y_1=0 \\
& Y_2=0 \\
& Y_3 =0.
\end{array} 
\right.  \label{r=0 in terms of Y}
\end{equation}
Thus, because of the presence of the term $\sqrt{\sigma}$, the condition $r=0$ is equivalent to (\ref{r=0 in terms of Y}), 
provided that $\sigma \geq 0$. If we now bear in mind that (\ref{g11})--(\ref{g34}) prevent $\sigma$ from vanishing, we can 
conclude that the only possible choice is $\sigma >0$, which is equivalent to (\ref{singularity kretschmann}). In other words, 
the presence of the 3-sphere where the Kretschmann invariant  is not defined follows directly from the condition $r=0$ which 
makes the curvature invariant (\ref{S-dS kretsch}) diverge. This fact can be interpreted as a hint indicating that this 3-sphere 
could represent a singularity of our ``boosted geometry''. Eventually, if we interpret $Y_0$ as the time coordinate 
(see (\ref{boost 1})), we can view 
(\ref{singularity kretschmann}) as a condition on time.

\begin{figure} [htbp] 
\includegraphics[scale=0.6]{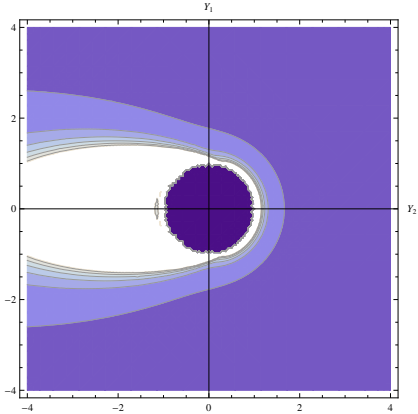}
\caption{Contour plot of the Kretschmann invariant numerically obtained with the following values of parameters: 
$a=1$, $m=0.1$, $Y_3 = Y_4 =0$ and $v=0.99$. The dark purple zone represents the circle of radius $a$ where the Kretschmann invariant is not defined.}
\label{contourplot}
\end{figure}

In the $Y_1-Y_2$ plane this 3-sphere becomes the circle with center at $Y_1 = Y_2 =0$ and radius $a$ of Fig. \ref{contourplot}, which 
represents a contour plot of the Kretschmann invariant, i.e. a plot where each different color corresponds to different 
values of the Kretschmann invariant and these values increase as we approach this circle.

\begin{figure} [htbp] 
\includegraphics[scale=0.6]{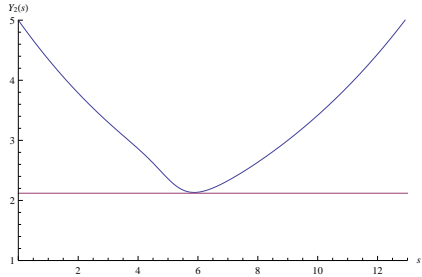}
\caption{Numerical solution of Eq. (\ref{geodesic eq}) for the function $Y_{2}(s)$ obtained in the $Y_1-Y_2$ plane and with initial 
conditions $Y_{1}(0)=Y_3(0)=Y_4(0)=0$, $Y_2(0)=5$, $Y^{\prime}_{1}(0)=Y^{\prime}_{3}(0)=Y^{\prime}_{4}(0)=0$ and 
$Y^{\prime}_{2}(0)=-0.7$. The values of parameters are $a=1$, $m=0.1$ and $v=0.9$. It is possible to see an ``antigravity effect'', 
since the function $Y_2(s)$ is pushed away from the ``boosted horizon'', which is represented by the horizontal line located at $Y_2=2.12$.}
\label{geod1}
\end{figure}

\begin{figure} [htbp] 
\includegraphics[scale=0.6]{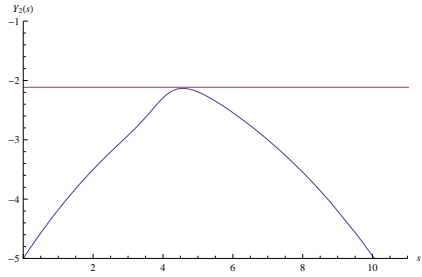}
\caption{Numerical solution of Eq. (\ref{geodesic eq}) for the function $Y_2(s)$ obtained in the $Y_1-Y_2$ plane and with initial 
conditions $Y_1(0)=Y_3(0)=Y_4(0)=0$, $Y_2(0)=-5$, $Y^{\prime}_{1}(0)=Y^{\prime}_{3}(0)=Y^{\prime}_{4}(0)=0$ and 
$Y^{\prime}_{2}(0)=0.9$. The values of parameters are $a=1$, $m=0.1$ and $v=0.9$.  The function $Y_2(s)$ initially moves toward the 
``boosted horizon'', i.e. the horizontal line at $Y_2=-2.12$, but then it is pushed away.}
\label{geod2}
\end{figure}

Another interesting feature of ``boosted geometries'' that we have found consists in the presence of a sort of barrier surrounding the  
3-sphere, which we may call ``boosted horizon'', in the sense that all geodesics, despite maintaining their completeness condition,  
are surprisingly pushed away from it.\footnote{More precisely, one defines an ``event horizon'' as the boundary of the causal past of future 
null infinity \cite{Haw-Ell}. In the ultrarelativistic regime we cannot say if this concept is still valid and hence we talk about 
``boosted horizon'' as the surface of spacetime surrounding the 3-sphere of radius $a$ where all geodesics, despite being 
complete, are pushed away.} We have also discovered that the extension of the ``boosted horizon'' depends on the boost velocity $v$, as we 
will shortly see. Since we have found that all geodesics are complete, according to standard definitions of general relativity the ``boosted horizon'' 
is not a singularity but, as we will show, it seems to be a sort of elastic wall which is hit by all particles before they get away.  
We have observed this effect numerically, by varying initial conditions of (\ref{geodesic eq}) and the boost velocity $v$, so as to 
reproduce different physical situations.  Figures \ref{geod1} and \ref{geod2} indeed represent one among the many situations 
analyzed which witness this ``antigravity'' effect. Figures \ref{geod1} and \ref{geod2}  show in fact a particle initially 
lying on the $Y_1=0$ line of Fig. \ref{contourplot} and having an 
initial velocity directed toward the region where the Kretschmann invariant is not defined. Strikingly, the solution ``refuses'' to be 
attracted by the 3-sphere but, regardless of its initial velocity, the particle always arrives at a certain point and then it goes away from it, 
as if an elastic wall were present. We propose to call this elastic wall ``boosted horizon''. The position of such a ``boosted horizon'' 
is independent of the initial velocity of the particle, but depends only on the boost velocity $v$. In fact, bearing in mind Fig. 
\ref{contourplot}, both for particles coming from ``above'' (i.e. particles initially lying on the positive half-line $Y_2>0$, 
$Y_1=0$ and with $Y^{\prime}_{2}(0) <0$) and for those coming from ``below'' (i.e. particles initially lying on the negative half-line 
$Y_2<0$, $Y_1=0$ and with $Y^{\prime}_{2}(0)>0$), the position of the ``boosted horizon'' does not change, as Tab. 
\ref{boosted horizon position} shows.
\begin{table}
\begin{tabular}{ |l | c|}
\hline
boost velocity &  ``boosted horizon'' location \\
& ($Y_2$ coordinate) \\
\hline
0.9995 & $\pm$ 1.02\\
0.9992 & $\pm$ 1.02\\
0.9991 & $\pm$ 1.02\\
0.999 & $\pm$ 1.02 \\
0.99  & $\pm$ 1.41 \\
0.9 & $\pm$ 2.12 \\
0.8 & $\pm$ 2.33 \\
0.7 & $\pm$ 2.43 \\
0.6 & $\pm$ 2.48 \\
0.5 & $\pm$ 2.48\\
0.4 & $\pm$ 2.42 \\
0.3 & $\pm$ 2.42 \\
0.2 & $\pm$ 2.34 \\
0.1 & $\pm$ 2.19 \\
0.01 & $\pm$ 1.52 \\
0.00155 & $\pm$ 1.00 \\
0.001 & $\pm$ 0.88 \\
0.0001 & $\pm$ 0.27 \\
\hline
\end{tabular}
\caption{\label{boosted horizon position} Location of the ``boosted horizon'' as a function of the boost velocity $v$. The positive 
sign refers to particles coming from ``above'' and the negative to those coming from ``below''. The values of parameters are $a=1$ and $m=0.1$.}
\end{table}
We have numerically checked, for each line of Tab. \ref{boosted horizon position}, 
that the minimum distance of the particle from the boundary of the $3$-sphere is always bigger than
its radius $a$, independently of the particle initial velocity. This means that the ``boosted horizon'' is always outside the
$3$-sphere. For example, we find that, when the boost velocity $v=0.5$, the minimum distance 
$d_{m}=3.1$ when $a=1$, and it decreases monotonically as $v$ increases or decreases, reaching a minimum value
of order $1.05 \div 1.10$. 

\begin{figure} [htbp] 
\includegraphics[scale=0.6]{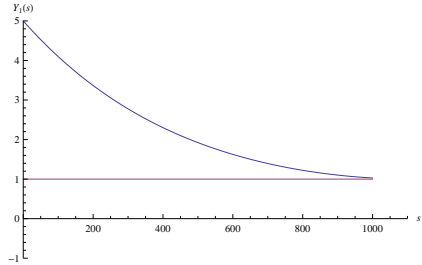}
\caption{Numerical solution of Eq. (\ref{geodesic eq}) for the function $Y_1(s)$ obtained in the $Y_1-Y_2$ plane and with initial 
conditions $Y_1(0)=5$, $Y_2(0)=Y_3(0)=Y_4(0)=0$, $Y^{\prime}_1(0)=-0.01$, $Y^{\prime}_2(0)=Y^{\prime}_3(0)=Y^{\prime}_4(0)=0$. 
The values of parameters are $a=1$, $m=0.1$ and $v=0.99$. The particle manages to hit the $3$-sphere, which is represented by the 
horizontal line $Y_1=1$.}
\label{geod3}
\end{figure}

The situation becomes somewhat intriguing when the particle lies initially on the $Y_2=0$ line (see Fig. \ref{contourplot}). 
In fact, in the cases in which the particle lies initially on the positive half-line $Y_1>0$, $Y_2=0$, it always manages to hit the 
3-sphere where the Kretschmann invariant is not defined, even if its initial velocity is extremely low, as we can see from Fig. 
\ref{geod3}. After the particle reaches the 3-sphere, its geodesic is not defined anymore and hence, according to definitions given above 
and those of Appendix \ref{b-completeness}, we can conclude that the 3-sphere of equation $(Y_{1})^2+(Y_{2})^2+(Y_{3})^2+(Y_{4})^2 = a^2$ defines 
a ``scalar curvature singularity'' for our ``boosted geometry'', as we have guessed before.

\begin{figure} [htbp] 
\includegraphics[scale=0.6]{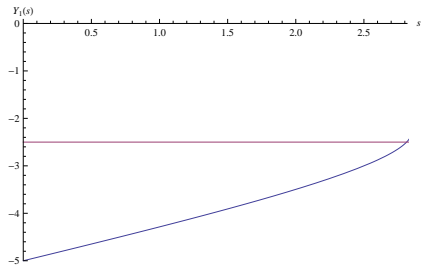}
\caption{Numerical solution of Eq. (\ref{geodesic eq}) for the function $Y_1(s)$ obtained in the $Y_1-Y_2$ plane and with initial 
conditions $Y_1(0)=-5$, $Y_2(0)=Y_3(0)=Y_4(0)=0$, $Y^{\prime}_1(0)=0.7$, $Y^{\prime}_2(0)=Y^{\prime}_3(0)=Y^{\prime}_4(0)=0$. 
The values of parameters are $a=1$, $m=0.1$ and $v=0.99$. The particle does not manage to hit the $3$-sphere but disappears in 
correspondence of the $Y_1=-2.5$ line. }
\label{geod4}
\end{figure}

\begin{figure} [htbp] 
\includegraphics[scale=0.6]{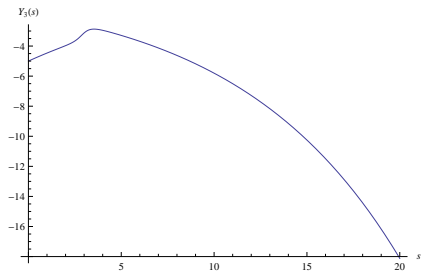}
\caption{Numerical solution of Eq. (\ref{geodesic eq}) for the function $Y_3(s)$ obtained in the $Y_3-Y_4$ plane and with initial 
conditions $Y_1(0)=Y_2(0)=0$, $Y_3(0)=Y_4(0)=-5$, $Y^{\prime}_1(0)=Y^{\prime}_2(0)=0$, $Y^{\prime}_3(0)
=Y^{\prime}_4(0)=0.566$. The values of parameters are $a=1$, $m=0.1$ and $v=0.99$. The ``antigravity effect'' is once again evident.}
\label{geod5}
\end{figure}

\begin{figure} [htbp] 
\includegraphics[scale=0.6]{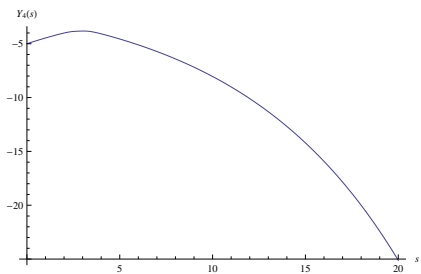}
\caption{Numerical solution of Eq. (\ref{geodesic eq}) for the function $Y_4(s)$ obtained in the $Y_3-Y_4$ plane and with initial 
conditions $Y_1(0)=Y_2(0)=0$, $Y_3(0)=Y_4(0)=-5$, $Y^{\prime}_1(0)=Y^{\prime}_2(0)=0$, $Y^{\prime}_3(0)
=Y^{\prime}_4(0)=0.566$. The values of parameters are $a=1$, $m=0.1$ and $v=0.99$. The ``antigravity effect'' is once again evident.}
\label{geod6}
\end{figure}

When the particle lies initially on the negative half-line $Y_1<0$, $Y_2=0$, its geodesic is not defined even before it reaches the 
$3$-sphere (see Fig. \ref{geod4}). This means that another ``scalar curvature singularity'' exists. Its position depends only on the boost 
velocity $v$ and not on the particle initial velocity. In any case, numerical analysis shows that this kind of 
singularities exists only if the particle lies initially on the $Y_2=0$ line.
We have repeated the same analysis also by putting $Y_1=Y_2=0$ in the relations defining the curvature, i.e. in the $Y_3-Y_4$ 
plane, and we have found the same ``antigravity effect'' of the previous cases, as shown in Figs. \ref{geod5} and \ref{geod6}, which 
represent some examples among the many situations numerically analyzed. Interestingly, in this case we have  found no 
``scalar curvature singularities''. 

In the ultrarelativistic regime ($v=0.9999$) the ``antigravity effects'' are still present but, as is clear from 
Tab. \ref{boosted horizon position}, the position of the boosted horizon tends to that of the singularity 3-sphere.

\section{The coordinate shift method}

A really important issue related to ``boosted geometries'' is represented by the occurrence of terms quadratic in Dirac's 
$\delta$ distribution in the Riemann tensor, which makes this object not defined. This Section has two purposes: on one hand 
it elucidates a new equivalent method for describing the gravitational field of a massless particle (showing that it gives the 
same results as the ones we have obtained in the previous Sections with the boosting procedure), on the other hand it proposes a 
recipe for the problem concerning the presence of products of two distributions in the Riemann tensor.  
   
The sources of gravitational (shock-) waves are massless particles moving along a null surface such as a horizon in the case 
of black holes. Therefore, another way to introduce a gravitational shock-wave is through a coordinate shift which reflects this 
peculiarity. This method is equivalent to the scissors-and-paste approach introduced by Penrose \cite{Penrose72} and can be applied 
both to vacuum solutions of Einstein equations \cite{Dray} and in presence of matter fields and non-vanishing 
cosmological constant \cite{Sfetsos95}.

Following Refs. \cite{Dray,Sfetsos95}, we start with the line element   
\begin{equation}
{\rm d}s^2 = 2 A(u,v) {\rm d}u {\rm d}v +g(u,v) h_{ij}(x) {\rm d}x^i{\rm d}x^j, \label{Sfetsosmetric}
\end{equation}
with $i,j=1,2$ (hereafter $v$ is a spacetime coordinate, unlike the previous Sections where it indicates the boost velocity 
or the particle velocity). We also assume the presence of some matter fields whose  non-vanishing components of 
the energy-momentum tensor are given by
\begin{equation}
T=2 \; T_{uv}(u,v,x)  \; {\rm d}u {\rm d}v + T_{uu}(u,v,x)  \; {\rm d}u^2 +  T_{vv}(u,v,x)  \; {\rm d}v^2 + T_{ij}(u,v,x)  \; {\rm d}x^i {\rm d}x^j.
\end{equation} 
 Consider a massless particle located at $u=0$ and moving with the speed of light in the $v$-direction. The coordinate 
shift method consists in making the ansatz according to which for $u<0$ the spacetime is still described by (\ref{Sfetsosmetric}) 
and for $u>0$ by (\ref{Sfetsosmetric}) but with $v$ shifted as $v \rightarrow v + f(x)$, where $f(x)$ is a (shift) function to 
be determined. Therefore, the resulting line element reads as
\begin{equation}
{\rm d}s^2 = 2 A(u,v+\Theta f) {\rm d}u\left( {\rm d}v +\Theta f_{,i} {\rm d}x^i  \right) +g(u,v+\Theta f) h_{ij}(x) 
{\rm d}x^i{\rm d}x^j, \label{shiftedmetric}
\end{equation} 
where $\Theta=\Theta(u)$ is the Heaviside step function and
\begin{equation}
\begin{split}
T=& 2 \; T_{uv}(u,v+\Theta f,x)  \; {\rm d}u ({\rm d}v+\Theta f_{,i} {\rm d}x^i )+ T_{uu}(u,v+\Theta f,x)  \; {\rm d}u^2  \\ 
&+ T_{vv}(u,v+\Theta f,x)  \; ( {\rm d}v+\Theta f_{,i} {\rm d}x^i )^2 + T_{ij}(u,v+\Theta f,x)  \; {\rm d}x^i {\rm d}x^j.
\end{split}
\end{equation} 
With the notation
\begin{equation}
\hat{u}=u, \;\;\;\;\; \hat{v}=v+f(x) \Theta(u), \;\;\;\;\; \hat{x}^i=x^i, 
\end{equation}
the metric (\ref{shiftedmetric}) assumes the handy form
\begin{equation}
\begin{split}
{\rm d}s^2 &= 2 \hat{A}\; {\rm d}\hat{u}\left( {\rm d}\hat{v} -\delta(\hat{u}) \hat{f} {\rm d} \hat{u}\right)
+ \hat{g}\; \hat{h}_{ij}(x) \;{\rm d}\hat{x}^i{\rm d}\hat{x}^j \\
&= 2 \hat{A}\; {\rm d}\hat{u} {\rm d}\hat{v} +\hat{F}\; {\rm d}\hat{u}^2 +\hat{g}\; \hat{h}_{ij}(x) 
\;{\rm d}\hat{x}^i{\rm d}\hat{x}^j, \label{handymetric}
 \end{split}
\end{equation}
and the energy-momentum tensor becomes
\begin{equation}
T= 2 \left(\hat{T}_{\hat{u}\hat{v}} -\hat{T}_{\hat{v}\hat{v}} \; \hat{f} \hat{\delta} \right)  {\rm d}\hat{u} {\rm d}\hat{v}  
+ \left(\hat{T}_{\hat{u}\hat{u}}+  \hat{T}_{\hat{v}\hat{v}} \; \hat{f}^2 \hat{\delta}^2 - 2 \hat{T}_{\hat{u}\hat{v}} 
\; \hat{f} \hat{\delta} \right) {\rm d}\hat{u}^2  
 + \hat{T}_{\hat{v}\hat{v}} {\rm d}\hat{v}^2 + \hat{T}_{ij} {\rm d}\hat{x}^i{\rm d}\hat{x}^j, \label{shifted_en-mom_tensor}
\end{equation}
with $\hat{F}=F(\hat{u},\hat{v},\hat{x})=-2 \; \hat{A} \; \hat{f} \; \hat{\delta}$ and where the hats indicate that the 
corresponding quantities are evaluated at $\hat{u},\hat{v},\hat{x}$ and $\hat{\delta}=\delta(\hat{u})$ is the $\delta$ distribution. 
We now demand that the metric (\ref{handymetric}) satisfies Einstein equation where the energy-momentum tensor is given by 
Eq. (\ref{shifted_en-mom_tensor}) plus the the energy-momentum tensor of the massless particle located at the origin of the 
transverse $x$-space and at $u=0$ and moving at the speed of light in the $v$-direction
\begin{equation}
T^p=T^p_{uu} {\rm d}u^2=\hat{T}^p_{\hat{uu}} {\rm d}\hat{u}^2=-4p\; \hat{A}^2 \hat{\delta}^{(2)}(\hat{x}) \hat{\delta}(\hat{u}){\rm d}\hat{u}^2,
\end{equation}
where $p$ is the particle momentum. If we suppose that the parts of field equations that do not involve the function $f$ are 
automatically satisfied, we find, by examining the terms linear in $f\; \delta$, that the necessary and sufficient conditions for 
being able to introduce a gravitational shock-wave via a coordinate shift amount to demand that at $u=0$ there exist the 
additional conditions (hereafter we drop the hat symbol to simplify the notation)
\begin{equation}
g_{,v}=A_{,v}=T_{vv}=0, \label{g,v-A,v}
\end{equation}
\begin{equation}
\triangle_{h_{ij}}f-\dfrac{g_{,uv}}{A}f=32 \pi \; p \; g \; A \; \delta^{(2)}(x), \label{f(x)_eq}
\end{equation}
where 
\begin{equation}
\triangle_{h_{ij}}=\dfrac{1}{\sqrt{h}} \partial_i \sqrt{h}h^{ij} \partial_j,
\end{equation}
is the Laplacian with respect to the 2-metric $h_{ij}$. A crucial point is represented by the presence of $\delta^2$ type 
terms both in Riemann and in Ricci tensors (see Appendix \ref{Riemann-Ricci}). These terms must vanish in a distributional sense, 
otherwise these two tensors are not defined. By considering the conditions (\ref{g,v-A,v}), it is easy to show that the quantities 
$\dfrac{A_{,vv}}{A}$, $\dfrac{A^2_{,v}}{A^2}$, $\dfrac{g_{,v}}{g}$, $\dfrac{A_{,v}}{A}$ appearing both in Riemann and in Ricci tensors 
are of order $O(u)$ or $O(u^2)$. Since all quantities involving $\delta$ terms should be intended as distributions to be integrated 
over smooth functions, we can conclude that all these $\delta^2$ terms give vanishing contribution and therefore both Riemann and 
Ricci tensors turn out to be under control as functions (in a distributional sense) of spacetime coordinates $\{u,v,x^1,x^2\}$, 
as advocated in Ref. \cite{Sfetsos95}.
The geodesic equations for the metric (\ref{handymetric}) obtained by varying the coordinates $v$ and $x^i$ are (dots denote 
derivatives with respect to the affine parameter)
\begin{equation}
\ddot{u}+ \dfrac{A_{,u}}{A} \dot{u}^2 -\dfrac{g_{,v}}{2 A} h_{ij} \dot{x}^i \dot{x}^j+f \; \dfrac{A_{,v}}{A} \delta \; \dot{u}^2=0,
\end{equation}
\begin{equation}
\ddot{x}^i + \Gamma^i_{\; jk} \dot{x}^j \dot{x}^k+\dfrac{g_{,u}}{g} \dot{u} \dot{x}^i +\dfrac{g_{,v}}{g} \dot{v} \dot{x}^i
+\dfrac{A}{g} \; \delta \; f_{,i} h^{ij} \dot{u}^2=0,
\end{equation}
where $\Gamma^i_{\; jk}$ denote the Christoffel symbols (see Appendix A of Ref. \cite{Sfetsos95} for their lengthy expression); 
the geodesic equation obtained from the variation of $u$ is 
\begin{equation}
\begin{split}
& \ddot{v}+ \dfrac{A_{,v}}{A} \dot{v}^2-\dfrac{g_{,u}}{2 A} h_{ij} \dot{x}^i \dot{x}^j + \left(   f \; 
\dfrac{A_{,u}}{A} \dot{u}^2 -2 f \; \dfrac{A_{,v}}{A} \dot{u} \dot{v}-2 f_{,i} \dot{u} \dot{x}^i   
-\dfrac{g_{,v}}{ A} \; f \; h_{ij} \dot{x}^i \dot{x}^j  \right) \delta  \\
& -f  \delta^{\prime} \dot{u}^2 + 2 f^2 \; \delta^2 \; \dfrac{A_{,v}}{A} \dot{u}^2 =0.
\end{split}
\end{equation}
On performing the integration of the geodesic equations, it is possible to understand how the original geometry 
(\ref{Sfetsosmetric}) is affected by the presence of a massless particle moving in the $v$-direction at $u=0$. 
In fact, as the geodesic trajectory crosses the null surface $u=0$ there is a shift in its $v$-component expressed by the relation
\begin{equation}
\Delta v \equiv v\vert_{u=0^+} - v\vert_{u=0^-}=f(x), \label{v_discontinuity}
\end{equation}
and a refraction effect in the transverse $x$-plane expressed by the refraction function
\begin{equation}
R^i(x) \equiv \dfrac{{\rm d}x^i}{{\rm d}u} \vert_{u=0^-} - \dfrac{{\rm d}x^i}{{\rm d}u} \vert_{u=0^+} 
= \frac{A}{g} \vert_{u=0} \; f_{,i} h^{ij}, \label{refraction_function}
\end{equation}
which measures the change of the angle that the trajectory forms with the $u=0$ surface after having crossed it. Therefore, 
when a trajectory crosses the $u=0$ null surface its $v$ component suffers from a discontinuity which, according to 
(\ref{v_discontinuity}), equals $f(x)$, while the other components remain continuous. Moreover, Eq. (\ref{refraction_function}) 
expresses the fact that the directional derivatives of $f(x)$ give information about how much the $x^i$ components change 
direction along $u$ while crossing the surface $u=0$. 

The examples analyzed in Ref. \cite{Sfetsos95} show that, in order to bring the metric (\ref{S-dS metric}) in the form 
(\ref{Sfetsosmetric}), we should introduce the function 
\begin{equation}
F:r \rightarrow  F(r)= {\rm exp} \left[ \dfrac{1}{a} \int {\rm d}r \; \dfrac{r a^2}{(r a^2-r^3-2 m a^2)} \right],
\end{equation} 
and the new independent variables
\begin{equation}
u= {\rm e}^{t/a} \; F(r), \; \; \; \; \; \; \; \; v={\rm e}^{-t/a} \; F(r). \label{coord_u,v}
\end{equation}
Bearing in mind these relation and Eq. (\ref{Sfetsosmetric}), we have that
\begin{eqnarray}
& A(u,v)=\dfrac{\left(1-{2m \over r}-{r^{2}\over a^{2}}\right)a^2}{2}\; F^{-2}, \label{A(u,v)} \\
& g(u,v)=r^2.
\end{eqnarray}
By performing the integration, we have found that
\begin{equation}
F(r)= {\rm exp} \left(   a\;   \dfrac{r_1(r_3-r_2)\log(r-r_1)+r_2(r_1-r_3)\log(r-r_2)
+r_3(r_2-r_1)\log(r-r_3)}{(r_1-r_2)(r_1-r_3)(r_2-r_3)}    \right), \label{F(r)_1}
\end{equation}
where $r_1$,$r_2$ and $r_3$ are the three roots of the cubic equation
\begin{equation}
r^3-ra^2+2ma^2=0, \label{cubic_eq}
\end{equation}
whose value is given by
\begin{equation}
r_1 = \dfrac{1}{3^{1/3}}\left(  \dfrac{a^2}{\Upsilon} + \dfrac{\Upsilon}{3^{1/3}} \right), \label{root_1} 
\end{equation}
\begin{equation}
r_{2,3} =\dfrac{1}{2}\dfrac{1}{3^{1/3}}\left(- \dfrac{\left(1 \pm i \sqrt{3} \; \right)a^2}{\Upsilon} 
- \dfrac{ \left(1\mp i \sqrt{3} \; \right) \Upsilon}{3^{1/3}} \right), \label{root_2and3}
\end{equation}
where $\Upsilon$ is defined as
\begin{equation}
\Upsilon \equiv \left( -9a^2m+\sqrt{3}\sqrt{27a^4m^2-a^6} \right)^{1/3}.
\end{equation}
In other words, Eqs. (\ref{root_1})--(\ref{root_2and3}) describe the three (null) surfaces where the metric (\ref{S-dS metric}) 
blows up, and hence the three horizons that characterize this geometry. With  the hypothesis $a/m >\sqrt{27}$ (which is 
respected by the choice $a=1$ and $m=0.1$ adopted in the last Section) the discriminant of (\ref{cubic_eq}) becomes negative 
and then (\ref{root_1})--(\ref{root_2and3}) turn out to be real roots. This condition allows us to write the roots 
(\ref{root_1})--(\ref{root_2and3}) in trigonometric form. We obtain 
\begin{equation}
 r_1= \dfrac{2a}{\sqrt{3}} \cos \left( \dfrac{\varphi}{3} \right), 
\end{equation}
\begin{equation}
r_{2,3} = -\dfrac{2a}{\sqrt{3}} \cos \left( \dfrac{\varphi \mp \pi}{3} \right)=-\dfrac{a}{\sqrt{3}} 
\left( \cos \dfrac{\varphi}{3} \pm \sqrt{3} \sin \dfrac{\varphi}{3} \right) \label{trigonometric_r3},
\end{equation}
where $\cos \varphi = \sqrt{27} m/a$. Note also that the roots (\ref{root_1})--(\ref{root_2and3}) are characterized 
by the fact that $r_1+r_2+r_3=0$ and $r_1r_2r_3=-2ma^2$.
Now we can write (\ref{F(r)_1}) as
\begin{equation}
F(r)=\prod_{i=1}^{3}(r-r_i)^{k_i}, \label{F(r)}
\end{equation}
where the three constants $k_i \; (i=1, \dots ,3)$ are given by 
\begin{equation}
 k_1=\dfrac{a r_1 (r_3-r_2)}{k_r} ,
\end{equation}
\begin{equation}
k_2=\dfrac{a r_2 (r_1-r_3)}{k_r} ,
\end{equation}
\begin{equation}
 k_3=\dfrac{a r_3 (r_2-r_1)}{k_r} ,
\end{equation}
with $k_r=(r_1-r_2)(r_1-r_3)(r_2-r_3)$. Therefore, bearing in mind (\ref{coord_u,v}) and (\ref{A(u,v)})  we have that
\begin{equation}
 A(u,v)=-\dfrac{1}{2r} \prod_{i=1}^{3}(r-r_i)^{1-2k_i}, 
\end{equation}
\begin{equation}
 u={\rm e}^{t/a} \prod_{i=1}^{3}(r-r_i)^{k_i}, 
\end{equation}
\begin{equation}
 v={\rm e}^{-t/a} \prod_{i=1}^{3}(r-r_i)^{k_i},
\end{equation}
and in particular we can satisfy the condition $u=0$ for $r=r_i \; (i=1, \dots ,3)$. Next, before writing down the equation 
satisfied by the shift function $f(\theta)$, we have to show that conditions (\ref{g,v-A,v}) are satisfied. Having obtained 
the following relations for the derivatives:
\begin{equation}
\dfrac{{\rm d}}{{\rm d}u}= \dfrac{1}{2} \left(\dfrac{a\;{\rm e}^{-t/a}}{F(r)}\dfrac{{\rm d}}{{\rm d}t}
+\dfrac{{\rm e}^{-t/a}}{F^\prime(r)}\dfrac{{\rm d}}{{\rm d}r} \right), 
\end{equation} 
\begin{equation}
\dfrac{{\rm d}}{{\rm d}v}= \dfrac{1}{2} \left(\dfrac{-a\;{\rm e}^{t/a}}{F(r)}\dfrac{{\rm d}}{{\rm d}t}
+\dfrac{{\rm e}^{t/a}}{F^\prime(r)}\dfrac{{\rm d}}{{\rm d}r} \right), 
\end{equation} 
we find that
\begin{equation}
g_{,v}= {\rm e}^{\frac{t}{a} } \; \frac{r (r-r_1)^{1-k_1} (r-r_2)^{1-k_2} (r-r_3)^{1-k_3}}{k_1 (r-r_2) (r-r_3)
+k_2(r-r_1) (r-r_3)+k_3 (r-r_1) (r-r_2)},
\end{equation}
and therefore
\begin{equation}
\lim_{u \rightarrow 0} g_{,v}=0  \; \; \; {\rm iff} \;  k_i <1. \label{g,v}
\end{equation}
Moreover, 
\begin{equation}
A_{,v}={\rm e}^{\frac{t}{a}} \; \frac{  (r-r_1)^{1-3 k_1} (r-r_2)^{1-3 k_2} (r-r_3)^{1-3k_3}} {4 r^2 \left[ k_1
(r-r_2) (r-r_3)+(r-r_1) (k_2 (r-r_3)+k_3 (r-r_2) \right]^2} \; \mathcal{F},
\end{equation}
where $\mathcal{F}=\mathcal{F}(r,r_i,k_i)$ is a function of $r$, the roots (\ref{root_1})--(\ref{root_2and3}) and the 
constants $k_i$ (which tend to a constant when $r \rightarrow r_i$), whose particular form is not of any special 
interest now. We can conclude that 
\begin{equation}
\lim_{u \rightarrow 0} A_{,v}=0  \; \; \; {\rm iff} \;  k_i <1/3. \label{A,v}
\end{equation}
By virtue of (\ref{g,v}) and (\ref{A,v}) we can say that conditions (\ref{g,v-A,v}) are satisfied provided that 
\begin{equation}
k_i < 1/3, \; \; (i=1, \dots ,3).
\end{equation}
By applying the shift coordinate method to the metric (\ref{S-dS metric}), we obtain the desired shock wave geometry and we 
find that the partial differential equation (\ref{f(x)_eq}) satisfied by the shift function $f(\theta)$ becomes  
\begin{equation}
\triangle_{(2)}f-c\; f=2 \pi k \delta(\xi-1) \delta (\phi), \label{f-SdS_eq}
\end{equation}
with
\begin{equation}
\triangle_{(2)}= \partial_\xi (1-\xi^2) \partial_\xi + \dfrac{\partial^2_\phi}{(1-\xi^2)}, \; \; \; \; \; \; 
{\rm with}\; \; \;  \xi=\cos \theta, 
\end{equation}
being the Laplacian on the unit 2-sphere having metric ${\rm d}s^2 ={\rm d}\theta^2 + \sin^2 \theta {\rm d} \phi^2$, 
and $k$ and $c$ being  real constants. This equation represents the usual Legendre equation of order $n$ ($n$ being a 
solution of $n(n+1)+c=0$) with a Dirac's $\delta$ appearing on the right-hand side. Therefore, its solutions depend strongly 
on the values assumed by the constant $c$ and can be given in terms of Legendre polynomials as
\begin{equation}
f(\theta;c)= - k \sum_{l=0}^{+ \infty} \dfrac{\left( l+\dfrac{1}{2}\right)}{\left[ l(l+1)+c\right]}P_l(\cos \theta), 
\; \; \; \; \; \;  c \in \mathbb{R}-\{-N(N+1), \; N=0,1,... \}. \label{f_solution}
\end{equation}  
In the case of Schwarzschild-de Sitter black hole, (\ref{f-SdS_eq}) depends on the ratio $a/m$ and thus possesses two branches 
of solutions for the constants $c$ and $k$. In the branch where the null surface is described by a positive value of $r$ we have that
\begin{equation}
c=\dfrac{(r_1-r_3)(r_3-r_2)}{a^2}=2 \sin \left(\dfrac{\varphi}{3} \right) \left[ \sqrt{3} \cos 
\left(\dfrac{\varphi}{3} \right) - \sin \left(\dfrac{\varphi}{3} \right) \right],
\end{equation}
while the constant $k$ is always positive, with precise value which is not of particular interest. The inequality 
$a/m >\sqrt{27}$ is equivalent to the obvious condition $\cos \varphi <1$, moreover the null hypersurface $u=0$ where the 
massless particle is placed corresponds to $r=r_3$ (see Eq. (\ref{trigonometric_r3})). The condition $r_3 > 0$ implies 
that (for positive values of $m$ and $a$ ) $\varphi \in (\pi/2, \dfrac{3}{2} \pi]$, so that
\begin{equation}
c \in (-2,0) \cup (0,1)  \; \; \; \; {\rm if} \; \; \; \varphi \in (\pi/2, \pi) \cup (\pi, \dfrac{3}{2} \pi ). 
\end{equation}
The boundary cases $c=-2$ ($\varphi = \dfrac{3}{2} \pi$) and $c=1$ ($\varphi = \pi/2$) correspond to de Sitter spacetime and 
Schwarzschild black hole, respectively, whereas the case $c=0$ ($\varphi= \pi$) is similar to the extremal Reissner-Nordstr\"{o}m 
charged black hole. The shift function $f(\theta)$ is given by Eq. (\ref{f_solution}). For $\dfrac{1}{4}\leq c <1$, an integral 
representation of the solution is given by \cite{Sfetsos95}
\begin{equation}
\begin{split}
f(\theta;c)&=\dfrac{-k}{\sqrt{2}} \int \limits_{0}^{+ \infty} {\rm d}s \; \cos (\sqrt{c-1/4} \;s) \dfrac{1}{\sqrt{\cosh s -\cos \theta}}  \\
&=\dfrac{- k \pi }{2 \cosh (\sqrt{c-1/4}\;\pi)} F (1/2-i\sqrt{c-1/4},1/2+i\sqrt{c-1/4};1;\cos^2 \dfrac{\theta}{2}),
\end{split}
\end{equation}
where $F(a,b;c;z)$ is the Gaussian or ordinary hypergeometric function. For $0 <c \leq \dfrac{1}{4}$ the solution is given by 
replacing $\sqrt{c-1/4}$ by $i\sqrt{1/4-c}$ and the trigonometric functions by hyperbolic ones, and vice versa. In both cases 
the shift function blows up at the point of the unit 2-sphere where the particle is located, i.e. at the northern pole $\theta=0$. 
Moreover, it is everywhere negative and for fixed $c$ it is a monotonically increasing function of $\theta \in [0, \pi]$, 
approaching a nonvanishing constant at $\theta=\pi$. For fixed $\theta$ it also monotonically increases as a function of $c \in (0,1)$. 
The refraction function (\ref{refraction_function}) is given by
\begin{equation}
R(\theta;c) = \left( \dfrac{A}{g} \right)_{u=0} \partial_{\theta} f(\theta;c).
\end{equation} 
It is a monotonically decreasing function of $\theta$ such that $\lim \limits_{\theta \to 0} R(\theta;c) = + \infty $ 
and $\lim \limits_{\theta \to \pi} R(\theta;c) = 0$. Thus, both the shift function and the refraction function blow up at 
$\theta=0$ and reach their minimum magnitudes at the southern pole $\theta=\pi$, where the refraction phenomenon disappears 
even if a particle trajectory is still discontinuous since $f(\pi;c) \neq 0$. For $-2<c<0$, the shift function is given 
by the integral representation 
\begin{equation}
f(\theta;c)=\dfrac{-k}{2c}-k \int \limits_{0}^{+ \infty} {\rm d}s \; \cosh (\sqrt{1/4-c} \;s) 
\left( \dfrac{1/\sqrt{2}}{\sqrt{\cosh s -\cos \theta}}-{\rm e}^{-s/2} \right). 
\end{equation}
The solution again blows up at $\theta=0$ and it monotonically increases as we move from $\theta=0$ to $\theta=\pi$. Moreover, 
it changes from negative to positive values at an angle $\theta_0$ that depends on the value assumed by the constant $c$ and 
reaches its minimum at $\theta=0$. On the other hand, the refraction function is a monotonically decreasing function of $\theta$.  

As we can see, the conditions found in this Section via the coordinate shift method are not in contrast with the results obtained 
through the boosting procedure of the previous Sections. We have shown in fact that the ``boosted horizon'' gives rise to a sort 
of ``antigravity effect'' which, in light of the results displayed in this Section, can be read as the refraction phenomenon 
described by the function (\ref{refraction_function}). It is an important result the fact these effects take place in nonsingular 
region of spacetime, i.e. the ``boosted horizon'' (for the boosting picture) and at the null hypersurface $u=0$ (in the coordinate 
shift method). Moreover, the presence of the singularity 3-sphere where the Kretschmann invariant is not defined could be probably 
related to the discontinuity of the $v$ component defined by Eq. (\ref{v_discontinuity}). The fact that in the ultrarelativistic 
regime the ``boosted horizon'' and the singularity 3-sphere positions' get blurred (as shown in Sec. III) represents a clue in 
favour of this hypothesis. To make more clear the equivalence between the boost and the coordinate shift method, one should be able 
to relate the $\{u,v,x^1,x^2 \}$ coordinates of the metric (\ref{handymetric}) with the $\{Y_1,Y_2,Y_3,Y_4 \}$ coordinates of the 
four-dimensional metric components (\ref{g11})--(\ref{g34}). This can be done with the help of the results of Appendix 
\ref{coordinate-transformation}. Therefore, by exploiting the equivalence between the two methods and the relations relating the 
two sets of coordinates, it is possible to relate all the results obtained through the coordinate shift method to those achieved 
with the boosting procedure. This means that also the considerations about how handling the $\delta^2$ terms in the Riemann tensor 
are valid also if we use the boost picture. Thus, the severe singularities of the Riemann tensor associated with metric 
(\ref{ultrarelativistic boosted metric}) can be considered to be under control. 

\section{Concluding remarks and open problems}

We have numerically evaluated, for the first time in the literature, the Riemann curvature of a boosted spacetime in the ultrarelativistic 
limit $v \rightarrow 1$,
starting from Schwarzschild-de Sitter spacetime metric (\ref{S-dS metric}). We have exploited the fact that a de Sitter spacetime can be 
seen as a four-dimensional hyperboloid embedded in a flat five-dimensional spacetime and satisfying the constraint 
(\ref{Z hyperboloid constraint}). After that, we have introduced the boosting procedure through the relations 
(\ref{boost 1})--(\ref{boost 2}) which make it possible to obtain the boosted Schwarzschild-de Sitter metric (\ref{boosted metric}), 
whose ultrarelativistic limit is represented by (\ref{ultrarelativistic boosted metric}). By exploiting the hyperboloid constraint 
(\ref{Z hyperboloid constraint}) we have then expressed (\ref{boosted metric}) in the manifestly four-dimensional form 
(\ref{g11})--(\ref{g34}). By virtue of (\ref{Z hyperboloid constraint}), the metric components (\ref{g11})--(\ref{g34}) are defined 
only if $\sigma >0$, $\sigma$ being defined by relation (\ref{Y0}). This fact is strictly related to inequality 
(\ref{singularity kretschmann}). In fact, $\left \{ \partial / \partial Y_\mu \right \}$ being a coordinate basis, we have numerically 
computed the Riemann curvature tensor by using the usual relations of general relativity, and to better understand the features of curvature 
we have studied both the Kretschmann invariant and the geodesic equation (\ref{geodesic eq}). We have indeed found that the Kretschmann invariant 
is not defined unless (\ref{singularity kretschmann}) holds and thus we have just concluded that there exists a 3-sphere of radius $a$ 
where the spacetime  possesses a ``scalar curvature singularity''. In fact, from the numerical analysis of the geodesic equation, we have 
found that if the particle lies initially on the positive half-line $Y_1>0$, $Y_2=0$ of Fig. \ref{contourplot} it  always reaches the 
3-sphere (Fig. \ref{geod3}). After that, its geodesic is no longer defined and hence we can conclude that the 3-sphere of equation 
$(Y_{1})^2+(Y_{2})^2+(Y_{3})^2+(Y_{4})^2 = a^2$ defines a ``scalar curvature singularity'' for the ``boosted geometry'' under investigation. When 
the particle lies initially on the negative half-line $Y_1<0$, $Y_2=0$, its geodesic is not defined even before it manages to reach the 
3-sphere (see Fig. \ref{geod4}): there exists a ``scalar curvature singularity'' whose position depends on the boost velocity $v$.

We have also discovered that ``boosted geometries'' are characterized by the presence of a sort of elastic wall surrounding the 3-sphere 
whose coordinates depend only on the boost velocity (see Tab. \ref{boosted horizon position}). All geodesics indeed, despite being complete, 
are always pushed away from there, as Figs. \ref{geod1} and \ref{geod2} show. We propose to call this barrier ``boosted horizon'' because, 
as in the case of Schwarzschild geometry, it is not a singularity of spacetime, but it is related to a sort of ``antigravity effect'' 
that should rule ``boosted geometries''. 

As we know, boosted geometries are ruled by the fact that both the spacetime metric and the Riemann curvature tensor assume a 
distributional nature in the ultrarelativistic regime. This regime is still ruled by ``antigravity effects'', with the peculiarity 
that ``boosted horizon'' and singularity 3-sphere tend to overlap. 

In the last part of the paper we have analyzed the geometry of the metric (\ref{S-dS metric}) through the coordinate shift method. 
We have shown that this new picture is equivalent to the boosting procedure and we have demonstrated how it solves the problem of the 
presence of $\delta^2$ terms in the Riemann tensor. In particular, the ``antigravity effects'' emerged at the ``boosted horizon'' 
have been ascribed to the refraction phenomenon described by the function (\ref{refraction_function}). Moreover, the fact that in the 
ultrarelativistic regime the ``boosted horizon'' position's tends to that of the singularity 3-sphere could be related to the fact that, 
in the coordinate shift method picture, when the particle crosses the null surface located at $u=0$ it suffers a discontinuity in its 
$v$-component (Eq. (\ref{v_discontinuity})) while the $x^i$ components are refracted according to (\ref{refraction_function}). This is 
a really delicate point as, unlike the singularity 3-sphere, both the null hypersurface $u=0$ (coordinate shift method) and our 
``boosted horizon'' (boosted picture) do not define a spacetime singularity, and we feel that some more efforts should be produced in 
this direction. The equivalence between the two methods (demonstrated in Ref. \cite{Dray}), which can be formally made manifest for our 
``boosted geometry'' by the results of Appendix \ref{coordinate-transformation}, has enabled us to conclude that the Riemann tensor 
associated with metric (\ref{ultrarelativistic boosted metric}) is defined and has a behavior under control.

We suppose that ``antigravity effects'' may result from the term $\Lambda = 3/a^2 >0$ occurring in the Schwarzschild-de Sitter metric 
(\ref{S-dS metric}) (a positive cosmological constant $\Lambda$ represents a repulsive interaction), while ``scalar curvature singularities'' 
might be related to the presence of a more exotic object, i.e. a firewall \cite{AMPS,Braunstein2013,Braunstein2015}, which can be 
a possible solution to an apparent 
inconsistency in black hole complementarity \cite{STU,SHW}.

\section*{acknowledgments} 
B.E. is grateful to Marcello Ortaggio for conversations. G. E. is grateful to the Dipartimento di Fisica of
Federico II University, Naples, for hospitality and support. 

\begin{appendix}

\section{The Riemann curvature tensor} \label{Appendix Riemann}

Since our paper is addressed to a wide physics audience, we recall here some basic properties of pseudo-Riemannian geometry. 
The Riemann tensor can be defined in various alternative (and equivalent) ways \cite{Wald}, \cite{Nakahara}. First, it can be defined 
as the map $R : \mathcal{X} (M)\otimes\mathcal{X} (M)\otimes\mathcal{X} (M) \rightarrow \mathcal{X} (M)$ ($\mathcal{X} (M)$ being the 
set of all vector fields defined on the manifold $M$) such that
\begin{equation}
R(X,Y,Z) \equiv \nabla_{X} \nabla_{Y} Z -\nabla_{Y} \nabla_{X} Z- \nabla_{[X,Y]}Z. \label{definition Riemann}
\end{equation}
In the case in which $[X,Y]=0$, the previous formula reduces to
\begin{equation}
R(X,Y,Z) \equiv \nabla_{X} \nabla_{Y} Z -\nabla_{Y} \nabla_{X} Z.
\end{equation}
Therefore, we can say that the Riemann tensor measures the failure of successive operations of differentiation to commute when applied 
to a dual vector field (which can be interpreted as the integrability obstruction for the existence of an isometry with Euclidean space), 
that is (in abstract index notation)
\begin{equation}
\nabla_a \nabla_b \; \omega_c - \nabla_b \nabla_a \; \omega_c =- R^{d}_{\; cab} \;  \omega_{d}.
\end{equation}
Moreover, we can say that the failure of a vector to return to its original value when parallel transported around a small closed loop 
is directly connected to the Riemann tensor, which is in this way related to the path dependence of parallel transport. We can easily 
construct a small closed loop at $p\in M$ by choosing a two-dimensional surface $S$ through $p$ and choosing coordinates $t$ and $s$ in 
the surface. Next we construct the loop by moving of a quantity $\Delta t$ along the $s=0$ curve, followed by moving $\Delta s$ along the 
$t= \Delta t$ curve and then we revert by $\Delta t$ and $\Delta s$. If we consider the vector $v^a$ at $p$ and parallel trasport it around 
the closed loop we have just constructed, the change $\delta v^a$ to second order in the displacement $\Delta t$, $\Delta s$ that we register 
when we revert to the starting point involves once again the Riemann tensor, because we have
\begin{equation}
\delta v^a = \Delta t \Delta s \; v^d \; T^c \; S^b \; R^{a}_{\; dcb},
\end{equation}
where $T^c$, $S^b$ indicate the tangent to the curves of constant $s$ and $t$, respectively. Finally, the Riemann tensor appears also in 
the geodesic deviation equation, the equation which measures the tendency of geodesics to accelerate toward or away from each other. 
If $\gamma_s (t)$ denotes a smooth $1$-parameter family of geodesics such that for each $s \in \mathbb{R}$ the curve $\gamma_s$ is a geodesic 
with affine parameter $t$, the geodesic deviation equation reads as
\begin{equation}
a^c \equiv T^a \nabla_a (T^b \nabla_b X^c) = R^{c}_{ \; def} T^d T^e X^f, \label{geodesic deviation}
\end{equation}
where $a^c$ is the relative acceleration of an infinitesimally nearby geodesic in the family,  $X^a = \partial x^a (s,t) /\partial s $ 
is the deviation vector ($x^a(s,t)$ being the coordinates of one geodesic of the family $\gamma_s (t)$) and  $ T^b 
=  \partial x^a (s,t) /\partial t$ represents the vector tangent to the geodesic. Therefore the equation (\ref{geodesic deviation}) 
states that, if the curvature does not vanish, some initially parallel geodesics will fail to remain parallel: in the presence of a 
gravitational field the fifth postulate of Euclidean geometry is no longer valid.

\section{The tetrad formalism} \label{Tetrad}

In most situations a curvature calculation that relies upon Christoffel symbols is extremely lengthy and not obviously feasible or 
readable. However, the tetrad
formalism is known to simplify such a task, at least when the metric does not possess distributional singularities. Thus, this appendix 
is devoted to some effort we made to express the highly singular ultrarelativistic boosted metric (\ref{ultrarelativistic boosted metric}) 
in terms of tetrads. 

As in the case of the boosted metric (\ref{boosted metric}), starting from the ultrarelativistic metric (\ref{ultrarelativistic boosted metric}) 
we can arrive at its manifestly four-dimensional form by exploiting (\ref{Y0}) and (\ref{dYo}) and hence we can eventually write the covariant 
metric components in the concise form
\begin{equation}
g_{kk}=1-{Y_{k}^{2}\over \sigma(Y_{j})}
+\left({Y_{k}^{2}\over \sigma(Y_{j})}+\delta_{1k}\right)
f(Y_{4})\delta\Bigr(Y_{1}+\sqrt{\sigma(Y_{j})}\Bigr), \;
\forall k=1,2,3,4,
\label{gkk ultrarelativ}
\end{equation}
\begin{equation}
g_{1k}=-{Y_{1}Y_{k}\over \sigma(Y_{j})}
+\left({Y_{1}\over \sqrt{\sigma(Y_{j})}}+1 \right)
{Y_{k}\over \sqrt{\sigma(Y_{j})}}f(Y_{4})
\delta \Bigr(Y_{1}+\sqrt{\sigma(Y_{j})}\Bigr), \;
\forall k=2,3,4,
\label{g1k ultrarelativ}
\end{equation}
\begin{equation}
g_{2k}=-{Y_{2}Y_{k}\over \sigma(Y_{j})}
+{Y_{2}Y_{k}\over \sigma(Y_{j})}f(Y_{4})
\delta \Bigr(Y_{1}+\sqrt{\sigma(Y_{j})}\Bigr), \;
\forall k=3,4,
\label{g2k ultrarelativ}
\end{equation}
\begin{equation}
g_{34}=-{Y_{3}Y_{4}\over \sigma(Y_{j})}
+{Y_{3}Y_{4}\over \sigma(Y_{j})}
f(Y_{4})\delta \Bigr(Y_{1}+\sqrt{\sigma(Y_{j})}\Bigr),
\label{g34 ultrarelativ}
\end{equation}
where
\begin{equation}
f(Y_{4}) \equiv 4p \left[-2+{Y_{4}\over a}
\log \left({{a+Y_{4}}\over {a-Y_{4}}}\right)\right].
\label{f(Y4)}
\end{equation}
Since all components of this metric are nonvanishing, at this
stage we still assume the existence of tetrad covectors $e_{\; \mu}^{a}$ such that the covariant form of the metric reads as
\begin{equation}
g_{\mu \nu}=e_{\; \mu}^{a} e_{\; \nu}^{b}\eta_{ab},
\label{metric tetrad}
\end{equation}
$a,b$ being Lorentz-frame indices, and $\eta_{ab}$ being the 
familiar Minkowski metric ${\rm diag}(-1,1,1,1)$. By comparison
of the formulae (\ref{gkk ultrarelativ})--(\ref{g34 ultrarelativ}) with (\ref{metric tetrad}) we find that one can set
\begin{equation}
e_{\; k}^{0}={Y_{k}\over \sqrt{\sigma(Y_{j})}}, \; 
\forall k=1,2,3,4,
\label{e0k}
\end{equation}
while the other components of the singular, distribution-valued
limit of tetrad covectors solve the following nonlinear 
algebraic system:
\begin{equation}
\Bigr(e_{\; k}^{1}\Bigr)^{2}+\Bigr(e_{\; k}^{2}\Bigr)^{2}
+\Bigr(e_{\; k}^{3}\Bigr)^{2}
=1+\left({Y_{k}^{2}\over \sigma(Y_{j})}+\delta_{1k} \right)
f(Y_{4})\delta \Bigr(Y_{1}+\sqrt{\sigma(Y_{j})}\Bigr), \;
\forall k=1,2,3,4,
\label{alg tetrad 1}
\end{equation}
\begin{equation}
\sum_{i=1}^{3}e_{\; 1}^{i} e_{\; k}^{i}
=\left({Y_{1}\over \sqrt{\sigma(Y_{j})}}+1 \right)
{Y_{k}\over \sqrt{\sigma(Y_{j})}}f(Y_{4})
\delta \Bigr(Y_{1}+\sqrt{\sigma(Y_{j})}\Bigr), \;
\forall k=2,3,4,
\label{alg tetrad 2}
\end{equation}
\begin{equation}
\sum_{i=1}^{3} e_{\; 2}^{i}e_{\; k}^{i}
={Y_{2}Y_{k}\over \sigma(Y_{j})}f(Y_{4})
\delta \Bigr(Y_{1}+\sqrt{\sigma(Y_{j})}\Bigr), \;
\forall k=3,4,
\label{alg tetrad 3}
\end{equation}
\begin{equation}
\sum_{i=1}^{3}e_{\; 3}^{i}e_{\; 4}^{i}
={Y_{3}Y_{4}\over \sigma(Y_{j})}
f(Y_{4})\delta \Bigr(Y_{1}+\sqrt{\sigma(Y_{j})}\Bigr).
\label{alg tetrad 4}
\end{equation}
Since the system  (\ref{alg tetrad 1})--(\ref{alg tetrad 4}) consists of $10$ equations for the $12$ unknown tetrad covectors, it is 
possible to find at least a particular solution. Now, once we get such a solution, the procedure should be as follows. As we know from 
general relativity, whenever the spacetime manifold
is parallelizable, we can always introduce a set of Lorentz frames \cite{Cartan}, so that the spin-connection $1$-form 
$\omega^{ab}=\omega_{\mu}^{ab}dx^{\mu}$ obtained from requiring
that the torsion $2$-form should vanish has components \cite{DW}
\begin{equation}
\omega_{\mu}^{ab}= {1\over 2}e^{a\nu}\Bigr(e_{\; \nu,\mu}^{b}
-e_{\; \mu,\nu}^{b}\Bigr)
-{1\over 2}e^{b \nu}\Bigr(e_{\; \nu,\mu}^{a}
-e_{\; \mu,\nu}^{a}\Bigr)
+{1\over 2}e^{a \nu}e^{b \sigma}\Bigr(e_{\; \nu,\sigma}^{c}
-e_{\; \sigma, \nu}^{c}\Bigr)e_{c \mu},
\label{spin connection}
\end{equation}
where
\begin{equation}
e^{a \nu}=\eta^{ab}e_{\; b}^{\nu}, \;
e_{c \mu}=e_{\; \mu}^{a}\eta_{ac},
\end{equation} 
the tetrad vectors $e_{\; a}^{\mu}$ being computable by
comparison from the relation
\begin{equation}
{\rm d}x^{\mu}=e_{\; a}^{\mu}e^{a},
\end{equation}
which holds by virtue of the definition of tetrad $1$-forms
\begin{equation}
e^{a} \equiv e_{\; \mu}^{a}dx^{\mu},
\end{equation}
jointly with \cite{DW}
\begin{equation}
e_{\; a}^{\rho}e_{\; \mu}^{a}=\delta_{\; \mu}^{\rho}.
\end{equation}
At this stage, we should be able to perform the curvature calculation
bearing in mind that the Riemann curvature is described by the $2$-form
\begin{equation}
R^{ab}={1\over 2}R_{\mu \nu}^{ab}dx^{\mu} \wedge dx^{\nu},
\label{curvature two form}
\end{equation}
where the components are given by
\begin{equation}
R_{\mu \nu}^{ab}=\Bigr(\omega_{\nu,\mu}^{ab}
-\omega_{\mu,\nu}^{ab}\Bigr)
+\eta_{cd}\Bigr(\omega_{\mu}^{bd}\omega_{\nu}^{ca}
-\omega_{\mu}^{ad}\omega_{\nu}^{cb}\Bigr).
\end{equation}
By virtue of Secs. II and III, the singular limit of the
curvature $2$-form is a nontrivial mathematical object, since
it involves the Dirac delta distribution, its fractional
powers and its derivatives. Finally, the Riemann curvature tensor 
$R_{\; \nu \rho \sigma}^{\mu}$ can be obtained from the identity
\begin{equation}
R_{\; \nu \rho \sigma}^{\mu} \; e_{\; \mu}^{a}
=R_{\; b \rho \sigma}^{a} \; e_{\; \nu}^{b}.
\end{equation}

\section{The b-completeness of spacetime}\label{b-completeness}

The b-boundary construction is a device to attach to any spacetime a set of boundary points. Such a boundary point can be considered as 
an equivalence class of inextendible curves in a spacetime, whose affine length is finite \cite{Haw-Ell,Sch}.

Let $\lambda(t)$ be a $C^1$ curve through a point $p$ of a manifold $M$ and let $\{ E_{\mu} \}$ (as before $\mu=1,2,3,4$) be a basis 
for the tangent vector space at $p$ to the manifold, $T_p M$. We can propagate $\{ E_{\mu} \}$ along $\lambda(t)$ to obtain a basis 
for $T_{\lambda (t)}M,  \; \forall t$. Then any $V = \left( \partial / \partial t \right)_{\lambda(t)} \in T_{\lambda (t)}M$ can be 
expressed as $V = V^\mu (t) E_\mu$ and we can define a generalized affine parameter $u$ on the curve $\lambda(t)$ by
\begin{equation}
u = \int_{p}  \left( \sum_{\mu} V_{\mu} V^{\mu} \right)^{1/2} {\rm d}t.
\end{equation}
Let $\{ E_{\mu^\prime} \}$ be another basis of $T_p M$. Then there exists some nonsingular matrix $A^{\mu}_{\; \; \nu}$ such that
\begin{equation}
E_{\nu} = \sum_{\mu^\prime} A^{\mu^{\prime}}_{\; \; \nu} E_{\mu^\prime}.
\end{equation}
As $\{ E_{\mu^\prime} \}$ and $\{ E_{\mu} \}$ are parallely transported along $\lambda(t)$, this relation is valid with constant 
$A^{\mu}_{\; \; \nu}$ and hence we have
\begin{equation}
V^{\mu^\prime} (t)= \sum_{\nu} A^{\mu^\prime}_{\; \; \nu} V^{\nu}(t).
\end{equation}
Since $A^{\mu}_{\; \; \nu}$ is nonsingular, there exists some constant $C>0$ such that
\begin{equation}
C \sum_{\mu} V_\mu V^\mu \leq \sum_{\mu^\prime} V_{\mu^\prime} V^{\mu^\prime} \leq C^{-1} \sum_{\mu} V_\mu V^\mu. 
\end{equation}
Thus, the length of a curve $\lambda$ is finite in the parameter $u$ if and only if it is finite in the parameter $u^\prime$. 
If $\lambda$ is a geodesic then $u$ becomes its affine parameter, but the definition given above is still valid since it has been formulated 
in terms of a general parameter $u$ defined on any $C^1$ curve. Therefore, we say that a spacetime $(M,g)$ is b-complete if there exists 
an endpoint for every $C^1$ curve of finite length as measured by a generalized affine parameter. We have that b-completeness implies 
g-completeness (short for geodesic completeness), but the converse is not true. Therefore, we can define a spacetime to be singularity-free 
if it is b-complete. This means that g-completeness represents the minimum condition for a spacetime to be considered singularity-free.

\section{$\delta^2$ terms in Riemann and Ricci tensors} \label{Riemann-Ricci}

We find that the only Riemann tensor components of the metric (\ref{handymetric}) depending on 
$\hat{\delta}^2=\delta^2(\hat{u})$ are given by (we drop the hat symbol to ease the notation)
\begin{equation}
R^{v}_{\; uvu}=2 \left(A_{,uv}-\dfrac{A_{,u}A_{,v}}{A} \right)f \delta + 2 \left(\dfrac{A_{,vv}}{A}-\dfrac{A^2_{,v}}{A^2} \right)f^2 \delta^2,
\end{equation} 
\begin{equation}
R^{v}_{\;ux^iu}=\left(2\dfrac{A_{,v}}{A}-\dfrac{g_{,v}}{g}\right)f_{,x^i}f\delta^2, \; \; \; \; (i=1,2),
\end{equation}
\begin{equation}
R^{x^i}_{\;ux^iu}=\left(\dfrac{g_{,v}}{g}\dfrac{A_{,v}}{A} \right)f^2 \delta^2 + 
\dots {\rm (terms \; at \; most \; linear \; in \; } \delta), \; \; \; \; (i=1,2).
\end{equation}
Therefore the only Ricci tensor component having $\delta^2$ terms is
\begin{equation}
\begin{split}
R_{uu} &=\sum_{\rho}R^{\rho}_{\; u\rho u}=R^{v}_{\; uvu}+ R^{x^1}_{\; ux^1u}+ R^{x^2}_{\; ux^2u} \\
&= 2 \left(\dfrac{A_{,vv}}{A}-\dfrac{A^2_{,v}}{A^2}+\dfrac{g_{,v}}{g}\dfrac{A_{,v}}{A} \right)f^2 \delta^2 
+ \dots {\rm (terms \; at \; most \; linear \; in \; } \delta).
\end{split}
\end{equation}

\section{Coordinate transformations} \label{coordinate-transformation}

In order to express the $\{\hat{u},\hat{v},\hat{\theta},\hat{\phi} \}$ coordinates characterizing the metric 
(\ref{handymetric}) in terms of $\{Y_1,Y_2,Y_3,Y_4 \}$ (which are the coordinates describing (\ref{g11})--(\ref{g34})), 
we start by inverting (\ref{boost 1})--(\ref{boost 2}), yielding easily
\begin{equation}
Y_0 = \gamma \left( Z_0 - v Z_1 \right),  \label{boost_Y0}
\end{equation}
 \begin{equation}
Y_1 = \gamma \left( Z_1-v Z_0  \right), 
\end{equation}
\begin{equation}
Y_2 = Z_2, \; \;  \; Y_3=Z_3 , \; \;  \; Y_4=Z_4. \label{boost_Y4}
\end{equation}
By using (\ref{Z0})--(\ref{Z3}) jointly with (\ref{boost_Y0})--(\ref{boost_Y4}) we obtain that
\begin{equation}
Y_0=\gamma \left( \sqrt{a^2-r^2} \sinh (t/a)-vr\cos \theta \right), 
\end{equation}
and
\begin{equation}
Y_1=\gamma \left( r\cos \theta-v\sqrt{a^2-r^2} \sinh (t/a) \right), \label{syst1}
\end{equation}
\begin{equation}
Y_2=r \sin \theta \cos \phi,\label{syst2}
\end{equation}
\begin{equation}
Y_3=r \sin \theta \sin \phi, \label{syst3}
\end{equation}
\begin{equation}
Y_4=\sqrt{a^2-r^2} \cosh(t/a).\label{syst4}
\end{equation}
Thus, bearing in mind that Eq. (\ref{hyperboloid constrain}) allows us to get rid of the $Y_0$ coordinate, 
if we want to obtain $\{t,r,\theta,\phi \}$ coordinates of Schwarzschild-de Sitter metric (\ref{S-dS metric}) 
as functions of $\{Y_1,Y_2,Y_3,Y_4 \}$ we have to invert relations (\ref{syst1})--(\ref{syst4}). First of all, 
by exploiting (\ref{boost 1})--(\ref{boost 2}), the condition $r^2=(Z_1)^2+(Z_2)^2+(Z_3)^2$ becomes
\begin{equation}
r^2=\gamma^2 (v \sqrt{\sigma}+Y_1)^2+(Y_2)^2+(Y_3)^2, \label{r^2_1}
\end{equation}
whereas on using (\ref{syst2}) and (\ref{syst3}) we obtain
\begin{equation}
r^2= \dfrac{(Y_2)^2+(Y_3)^2}{\sin^2 \theta},\label{r^2_2}
\end{equation}
therefore a comparison of (\ref{r^2_1}) and (\ref{r^2_2}) yields
\begin{equation}
\sin^2 \theta = \dfrac{(Y_2)^2+(Y_3)^2}{\gamma^2 (v \sqrt{\sigma}+Y_1)^2+(Y_2)^2+(Y_3)^2},
\end{equation}
whose solutions are given by
\begin{equation}
\theta = \mp \arcsin \left(   \sqrt{\dfrac{(Y_2)^2+(Y_3)^2}{\gamma^2 (v \sqrt{\sigma}+Y_1)^2+(Y_2)^2+(Y_3)^2}} \right) 
+ 2\pi n, \;\; \; (n\; {\rm integer}), \label{theta-Y_1}
\end{equation}
\begin{equation}
\theta = \pi \mp \arcsin \left(   \sqrt{\dfrac{(Y_2)^2+(Y_3)^2}{\gamma^2 (v \sqrt{\sigma}+Y_1)^2+(Y_2)^2+(Y_3)^2}} 
\right) + 2\pi n, \;\; \; (n\; {\rm integer}), \label{theta-Y_2}
\end{equation}
therefore at this stage from (\ref{syst4}) we straightforwardly obtain the relations for $t$, i.e. 
\begin{equation}
t= a \left[ \mp {\rm arccosh}\left(\dfrac{Y_4}{\sqrt{a^2-r^2}} \right) +2 \pi i n   \right], \; \;\; (n\; {\rm integer}), \label{t-Y}
\end{equation}
and eventually from (\ref{syst3}) we get 
\begin{equation}
\phi = \arcsin \left( \dfrac{Y_3}{r \sin \theta} \right) + 2 \pi n, \; \;\; (n\; {\rm integer}), \label{phi-Y_1}
\end{equation}
\begin{equation}
\phi = \pi-\arcsin \left( \dfrac{Y_3}{r \sin \theta} \right) + 2 \pi n, \; \;\; (n\; {\rm integer}). \label{phi-Y_2}
\end{equation}
Thus, Eqs. (\ref{r^2_1}), (\ref{theta-Y_1}) and (\ref{theta-Y_2}), (\ref{t-Y}), (\ref{phi-Y_1}) and (\ref{phi-Y_2}), 
represent the relations which link $\{t,r,\theta,\phi \}$ to $\{Y_1,Y_2,Y_3,Y_4 \}$ coordinates. By exploiting these 
relations, it is possible to express the $\{\hat{u},\hat{v},\hat{\theta},\hat{\phi} \}$ coordinates as functions 
of $\{Y_1,Y_2,Y_3,Y_4 \}$ and to define the equivalence between the boosting procedure and the coordinate shift method.

\end{appendix}

\end{document}